\documentclass[aps,prl,preprint,tightenlines,superscriptaddress,showpacs,byrevtex]{revtex4}

\usepackage{epsfig,floatflt}
\usepackage{graphicx} 
\usepackage{dcolumn}  
\usepackage{amssymb,amsmath,indentfirst,enumerate}

\graphicspath{{ps}}

\newcommand{\UFS}{\ensuremath{\Upsilon(10860)}}
\newcommand{\USS}{\ensuremath{\Upsilon(11020)}}

\newcommand{\pp}{\ensuremath{\pi^+\pi^-}}
\newcommand{\uu}{\ensuremath{\mu^+\mu^-}}
\newcommand{\ee}{\ensuremath{e^+e^-}}

\begin{document}

\preprint{\vbox{ \hbox{   }
                 \hbox{BELLE-CONF-1605}
}}

\title{\quad\\[0.5cm]
Study of Two-Body {\boldmath $\ee\to B_s^{(*)}\bar{B}_s^{(*)}$} 
Production in the 
Energy Range from 10.77 to 11.02 GeV}


\noaffiliation
\affiliation{Aligarh Muslim University, Aligarh 202002}
\affiliation{University of the Basque Country UPV/EHU, 48080 Bilbao}
\affiliation{Beihang University, Beijing 100191}
\affiliation{University of Bonn, 53115 Bonn}
\affiliation{Budker Institute of Nuclear Physics SB RAS, Novosibirsk 630090}
\affiliation{Faculty of Mathematics and Physics, Charles University, 121 16 Prague}
\affiliation{Chiba University, Chiba 263-8522}
\affiliation{Chonnam National University, Kwangju 660-701}
\affiliation{University of Cincinnati, Cincinnati, Ohio 45221}
\affiliation{Deutsches Elektronen--Synchrotron, 22607 Hamburg}
\affiliation{University of Florida, Gainesville, Florida 32611}
\affiliation{Department of Physics, Fu Jen Catholic University, Taipei 24205}
\affiliation{Justus-Liebig-Universit\"at Gie\ss{}en, 35392 Gie\ss{}en}
\affiliation{Gifu University, Gifu 501-1193}
\affiliation{II. Physikalisches Institut, Georg-August-Universit\"at G\"ottingen, 37073 G\"ottingen}
\affiliation{SOKENDAI (The Graduate University for Advanced Studies), Hayama 240-0193}
\affiliation{Gyeongsang National University, Chinju 660-701}
\affiliation{Hanyang University, Seoul 133-791}
\affiliation{University of Hawaii, Honolulu, Hawaii 96822}
\affiliation{High Energy Accelerator Research Organization (KEK), Tsukuba 305-0801}
\affiliation{J-PARC Branch, KEK Theory Center, High Energy Accelerator Research Organization (KEK), Tsukuba 305-0801}
\affiliation{Hiroshima Institute of Technology, Hiroshima 731-5193}
\affiliation{IKERBASQUE, Basque Foundation for Science, 48013 Bilbao}
\affiliation{University of Illinois at Urbana-Champaign, Urbana, Illinois 61801}
\affiliation{Indian Institute of Science Education and Research Mohali, SAS Nagar, 140306}
\affiliation{Indian Institute of Technology Bhubaneswar, Satya Nagar 751007}
\affiliation{Indian Institute of Technology Guwahati, Assam 781039}
\affiliation{Indian Institute of Technology Madras, Chennai 600036}
\affiliation{Indiana University, Bloomington, Indiana 47408}
\affiliation{Institute of High Energy Physics, Chinese Academy of Sciences, Beijing 100049}
\affiliation{Institute of High Energy Physics, Vienna 1050}
\affiliation{Institute for High Energy Physics, Protvino 142281}
\affiliation{Institute of Mathematical Sciences, Chennai 600113}
\affiliation{INFN - Sezione di Torino, 10125 Torino}
\affiliation{Advanced Science Research Center, Japan Atomic Energy Agency, Naka 319-1195}
\affiliation{J. Stefan Institute, 1000 Ljubljana}
\affiliation{Kanagawa University, Yokohama 221-8686}
\affiliation{Institut f\"ur Experimentelle Kernphysik, Karlsruher Institut f\"ur Technologie, 76131 Karlsruhe}
\affiliation{Kavli Institute for the Physics and Mathematics of the Universe (WPI), University of Tokyo, Kashiwa 277-8583}
\affiliation{Kennesaw State University, Kennesaw, Georgia 30144}
\affiliation{King Abdulaziz City for Science and Technology, Riyadh 11442}
\affiliation{Department of Physics, Faculty of Science, King Abdulaziz University, Jeddah 21589}
\affiliation{Korea Institute of Science and Technology Information, Daejeon 305-806}
\affiliation{Korea University, Seoul 136-713}
\affiliation{Kyoto University, Kyoto 606-8502}
\affiliation{Kyungpook National University, Daegu 702-701}
\affiliation{\'Ecole Polytechnique F\'ed\'erale de Lausanne (EPFL), Lausanne 1015}
\affiliation{P.N. Lebedev Physical Institute of the Russian Academy of Sciences, Moscow 119991}
\affiliation{Faculty of Mathematics and Physics, University of Ljubljana, 1000 Ljubljana}
\affiliation{Ludwig Maximilians University, 80539 Munich}
\affiliation{Luther College, Decorah, Iowa 52101}
\affiliation{University of Maribor, 2000 Maribor}
\affiliation{Max-Planck-Institut f\"ur Physik, 80805 M\"unchen}
\affiliation{School of Physics, University of Melbourne, Victoria 3010}
\affiliation{Middle East Technical University, 06531 Ankara}
\affiliation{University of Miyazaki, Miyazaki 889-2192}
\affiliation{Moscow Physical Engineering Institute, Moscow 115409}
\affiliation{Moscow Institute of Physics and Technology, Moscow Region 141700}
\affiliation{Graduate School of Science, Nagoya University, Nagoya 464-8602}
\affiliation{Kobayashi-Maskawa Institute, Nagoya University, Nagoya 464-8602}
\affiliation{Nara University of Education, Nara 630-8528}
\affiliation{Nara Women's University, Nara 630-8506}
\affiliation{National Central University, Chung-li 32054}
\affiliation{National United University, Miao Li 36003}
\affiliation{Department of Physics, National Taiwan University, Taipei 10617}
\affiliation{H. Niewodniczanski Institute of Nuclear Physics, Krakow 31-342}
\affiliation{Nippon Dental University, Niigata 951-8580}
\affiliation{Niigata University, Niigata 950-2181}
\affiliation{University of Nova Gorica, 5000 Nova Gorica}
\affiliation{Novosibirsk State University, Novosibirsk 630090}
\affiliation{Osaka City University, Osaka 558-8585}
\affiliation{Osaka University, Osaka 565-0871}
\affiliation{Pacific Northwest National Laboratory, Richland, Washington 99352}
\affiliation{Panjab University, Chandigarh 160014}
\affiliation{Peking University, Beijing 100871}
\affiliation{University of Pittsburgh, Pittsburgh, Pennsylvania 15260}
\affiliation{Punjab Agricultural University, Ludhiana 141004}
\affiliation{Research Center for Electron Photon Science, Tohoku University, Sendai 980-8578}
\affiliation{Research Center for Nuclear Physics, Osaka University, Osaka 567-0047}
\affiliation{Theoretical Research Division, Nishina Center, RIKEN, Saitama 351-0198}
\affiliation{RIKEN BNL Research Center, Upton, New York 11973}
\affiliation{Saga University, Saga 840-8502}
\affiliation{University of Science and Technology of China, Hefei 230026}
\affiliation{Seoul National University, Seoul 151-742}
\affiliation{Shinshu University, Nagano 390-8621}
\affiliation{Showa Pharmaceutical University, Tokyo 194-8543}
\affiliation{Soongsil University, Seoul 156-743}
\affiliation{University of South Carolina, Columbia, South Carolina 29208}
\affiliation{Stefan Meyer Institute for Subatomic Physics, Vienna 1090}
\affiliation{Sungkyunkwan University, Suwon 440-746}
\affiliation{School of Physics, University of Sydney, New South Wales 2006}
\affiliation{Department of Physics, Faculty of Science, University of Tabuk, Tabuk 71451}
\affiliation{Tata Institute of Fundamental Research, Mumbai 400005}
\affiliation{Excellence Cluster Universe, Technische Universit\"at M\"unchen, 85748 Garching}
\affiliation{Department of Physics, Technische Universit\"at M\"unchen, 85748 Garching}
\affiliation{Toho University, Funabashi 274-8510}
\affiliation{Tohoku Gakuin University, Tagajo 985-8537}
\affiliation{Department of Physics, Tohoku University, Sendai 980-8578}
\affiliation{Earthquake Research Institute, University of Tokyo, Tokyo 113-0032}
\affiliation{Department of Physics, University of Tokyo, Tokyo 113-0033}
\affiliation{Tokyo Institute of Technology, Tokyo 152-8550}
\affiliation{Tokyo Metropolitan University, Tokyo 192-0397}
\affiliation{Tokyo University of Agriculture and Technology, Tokyo 184-8588}
\affiliation{University of Torino, 10124 Torino}
\affiliation{Toyama National College of Maritime Technology, Toyama 933-0293}
\affiliation{Utkal University, Bhubaneswar 751004}
\affiliation{Virginia Polytechnic Institute and State University, Blacksburg, Virginia 24061}
\affiliation{Wayne State University, Detroit, Michigan 48202}
\affiliation{Yamagata University, Yamagata 990-8560}
\affiliation{Yonsei University, Seoul 120-749}
  \author{A.~Abdesselam}\affiliation{Department of Physics, Faculty of Science, University of Tabuk, Tabuk 71451} 
  \author{I.~Adachi}\affiliation{High Energy Accelerator Research Organization (KEK), Tsukuba 305-0801}\affiliation{SOKENDAI (The Graduate University for Advanced Studies), Hayama 240-0193} 
  \author{K.~Adamczyk}\affiliation{H. Niewodniczanski Institute of Nuclear Physics, Krakow 31-342} 
  \author{H.~Aihara}\affiliation{Department of Physics, University of Tokyo, Tokyo 113-0033} 
  \author{S.~Al~Said}\affiliation{Department of Physics, Faculty of Science, University of Tabuk, Tabuk 71451}\affiliation{Department of Physics, Faculty of Science, King Abdulaziz University, Jeddah 21589} 
  \author{K.~Arinstein}\affiliation{Budker Institute of Nuclear Physics SB RAS, Novosibirsk 630090}\affiliation{Novosibirsk State University, Novosibirsk 630090} 
  \author{Y.~Arita}\affiliation{Graduate School of Science, Nagoya University, Nagoya 464-8602} 
  \author{D.~M.~Asner}\affiliation{Pacific Northwest National Laboratory, Richland, Washington 99352} 
  \author{T.~Aso}\affiliation{Toyama National College of Maritime Technology, Toyama 933-0293} 
  \author{H.~Atmacan}\affiliation{Middle East Technical University, 06531 Ankara} 
% ...
\collaboration{The Belle Collaboration}



\begin{abstract}

We report results on the studies of the $\ee\to B_s^{(*)}\bar{B}_s^{(*)}$ 
processes.
The results are based on a $121.4$~fb$^{-1}$ data sample collected with the 
Belle detector at the center-of-mass energy near the $\UFS$ peak and
$16.4$~fb$^{-1}$ of data collected at 19 energy points in the range from 
10.77 to 11.02~GeV. We observe a clear 
$\ee\to\UFS\to B_s^{(*)}\bar{B}_s^{(*)}$ signal, with no 
statistically significant signal of 
$\ee\to\Upsilon(11020)\to B_s^{(*)}\bar{B}_s^{(*)}$.
The relative production ratio of $B_s^*\bar{B}_s^*$, $B_s\bar{B}_s^{*}$,
and $B_s\bar{B}_s$ final states at $\sqrt{s}=10.866$~GeV is measured
to be 
$7:$ $0.856\pm0.106(stat.)\pm0.053(syst.):$ 
$0.645\pm0.094(stat.)^{+0.030}_{-0.033}(syst.)$.
An angular analysis of the $B_s^*\bar{B}_s^*$ final state produced at the 
$\UFS$ peak is also performed.

\end{abstract}

\maketitle


\section{Introduction}
The Belle experiment has recently measured the ratio
$R_b=\sigma_{\ee\to b\bar{b}}/\sigma_{\ee\to\uu}$ in the energy range 
from 10.60 to 11.02~GeV utilizing an inclusive technique~\cite{ee2bb}. 
In addition, the energy dependence of the production cross section has been 
studied for several exclusive channels such as $\ee\to\Upsilon(nS)\pp$ 
($n=1,2,3$)~\cite{ee2bb} and $\ee\to h_b(mP)\pp$ ($m=1,2$)~\cite{hpp-cs}.
The measured energy dependence for the aforementioned exclusive cross 
sections exhibits substantially different behaviour compared to that for 
$R_b$. Measurements of the cross sections for other exclusive final states, 
such as two-body $B^{(*)}\bar{B}^{(*)}$, $B_s^{(*)}\bar{B}_s^{(*)}$,
and three-body $B^{(*)}\bar{B}^{(*)}\pi$, might shed light on the mechanisms 
of the $b\bar{b}$ hadronization and on the nature of the $\UFS$ and 
$\USS$ resonances.

In this paper, we present preliminary results on the analysis of the 
$\ee\to B_s^{(*)}\bar{B}_s^{(*)}$ processes in the energy range from 10.77
to 11.02 GeV in the center-of-mass (c.m.) frame using data accumulated 
with the Belle detector~\cite{Belle} operating at the asymmetric-energy 
$\ee$ collider KEKB~\cite{KEKB}.


\section{The Belle detector}

The Belle detector is a large-solid-angle magnetic spectrometer
based on a 1.5~T superconducting solenoid magnet. Charged particle tracking is
provided by a four-layer silicon vertex detector and a 50-layer central
drift chamber (CDC) that surround the interaction point. The charged particle
acceptance covers laboratory polar angle $\theta$ between $17^{\circ}$ and
$150^{\circ}$, corresponding to about 92\% of the total solid angle in the
c.m.\ frame. 

Charged hadron identification is provided by $dE/dx$ measurements in the CDC,
an array of 1188 aerogel Cherenkov counters (ACC), and a barrel-like array
of 128 time-of-flight scintillation counters (TOF); information from the three
subdetectors is combined to form a single likelihood ratio, which is then used
in kaon and pion selection. Electromagnetic showering particles are
detected in an array of 8736 CsI(Tl) crystals (ECL) that covers nearly the 
same solid angle as the charged particle tracking system. 

Electron identification in Belle is based on a combination of $dE/dx$
measurements in the CDC, the response of the ACC, and the position, shape 
and total energy deposition of the shower detected in the ECL.
The electron identification efficiency is greater than 92\% for tracks with
$p_{\rm lab}>1.0$~GeV/$c$ and the hadron misidentification probability is 
below 0.3\%. The magnetic field is returned via an iron yoke that is 
instrumented to detect muons and $K^0_L$ mesons. Muons are identified based
on their penetration range and transverse scattering in this KLM detector. 
In the momentum region relevant to this analysis, the identification 
efficiency is about 90\% while the probability to misidentify a pion as a 
muon is below 2\%. 

We use the EvtGen event generator~\cite{EvtGen} with PHOTOS~\cite{PHOTOS} 
for radiative corrections and a GEANT-based Monte 
Carlo (MC) simulation~\cite{GEANT} to model the response of the detector
and determine the acceptance. The MC simulation includes run-dependent 
detector performance variations and background conditions.



\section{Event Reconstruction}

Charged tracks are selected with a set of track quality requirements based 
on the number of CDC hits and on the distances of closest approach to the
interaction point (IP) along (perpendicular to) the beam axis of $|dz|<5$~cm 
($(dr)<2.5$~cm). Tracks originating from a $B_s$ candidate are required 
to have momenta transverse to the beam greater than 0.05~GeV/$c$. 
For charged kaon identification, we impose a particle-identification 
requirement that has an 86\% efficiency and a 7\% fake rate from 
misidentified pions. Charged hadron candidates that are positively 
identified as electrons are excluded. 


\subsection{$B_s$ Reconstruction}

Candidate $B_s$ decays are reconstructed in the following channels:
$B_s\to D_s^{(*)-}\pi^+$, $B_s\to J/\psi K^+K^-$, $B_s\to J/\psi \pi^+\pi^-$, 
and $B_s\to\psi(2S) K^+K^-$. Candidate $D^*_s$ decays are reconstructed 
in the $D_s\gamma$ channel, where $D_s\to K^+K^-\pi^-$ or $K^0_SK^-$. 
$D_s$ candidates from the $B_s\to D_s^-\pi^+$ decay mode are reconstructed 
in the $K^+K^-\pi^-$, $K^0_SK^-$, and $K^0_SK^+\pi^-\pi^-$ final states. 
Neutral kaon ($K^0_S$) candidates are reconstructed using pairs of 
oppositely-charged tracks, both treated as pions, with an invariant mass 
within 15~MeV/$c^2$ of the nominal $K^0_S$ mass; the IP constraint is not 
imposed here. The direction of the $K^0_S$ candidate momentum vector is 
required to be consistent with the direction of its vertex displacement 
relative to the IP. To identify 
signal $D_s$ [$D^*_s$] candidates, we require $|M(D_s)-m_{D_s}|<2.5\sigma$ 
[$|\big(M(D_s\gamma)-M(D_s)\big)-\big(m_{D^*_s}-m_{D_s}\big)|<2.5\sigma$], 
where $m_{D_s}$ [$m_{D^*_s}$] is the $D_s$ [$D^*_s$] nominal 
mass~\cite{PDG}, and $\sigma$ is the Gaussian width for the relevant
final state. The invariant mass of the $J/\psi\to\ell^+\ell^-$ candidates,
with $\ell$ being electron (muon), is required to satisfy 
$3.01~(3.05)$~GeV/$c^2<M(\ell^+\ell^-)<3.13$~GeV/$c^2$. The $\psi(2S)$ 
candidates are reconstructed in the $\psi(2S)\to J/\psi\pi^+\pi^-$ decay 
mode. We require 
$|\big(M(J/\psi\pi^+\pi^-)-M(J/\psi)\big)-
  \big(m_{\psi(2S)}-m_{J/\psi}\big)|<8$~MeV/$c^2$,
where $m_{J/\psi}$ and $m_{\psi(2S)}$ are the $J/\psi$ and $\psi(2S)$ nominal
masses~\cite{PDG}, respectively. 

We identify $B_s$ candidates by their reconstructed invariant mass 
$M(B_s)$ and momentum $P(B_s)$. We do not reconstruct the photon from the 
$B^*_s\to B_s\gamma$ decay; instead, the individual two-body final states 
are discriminated based on the reconstructed $B_s$ momentum. Signal 
$\UFS\to B_s^*\bar{B}_s^*$ events produce a narrow peak in the $P(B_s)$ 
spectrum around $0.442$~GeV/$c$, the $\UFS\to B_s\bar{B}_s^*$ signal events 
produce a peak at $0.678$~GeV/$c$, and $\UFS\to B_s\bar{B}_s$ signal peaks 
at $0.844$~GeV/$c$. It is important to note here that, due to the very low 
momentum of the photon from the $B_s^*\to B_s\gamma$ decays, the 
$B_s\bar{B}_s^*$ events (where the reconstructed $B_s$ is the one from 
$B_s^*$) produce a peak in the $P(B_s)$ distribution at about the same 
position as $B_s\bar{B}_s^*$ events, where the reconstructed $B_s$ is the 
prompt one. This is confirmed with the signal MC simulation. Momentum 
smearing for $B_s$ daughters from $B^*_s$ decays becomes more significant 
for higher $E_{\rm cm}$ values. 


\subsection{Background Suppression}

The dominant source of background arises from $e^+e^-\to c\bar{c}$ continuum
events, where real $D$ mesons produced in $e^+e^-$ annihilation are combined
with random particles to form a $B$ candidate. This type of background is
suppressed using variables that characterize the event topology. Since the
momenta of the $B_s^{(*)}$ and $\bar{B}_s^{(*)}$ mesons produced from the 
$\UFS$ decay are low in the c.m.\ frame, their decay products are 
essentially uncorrelated 
and the event tends to be spherical. In contrast, hadrons from continuum 
events tend to exhibit a two-jet structure. We use $\theta_{\rm thr}$, 
the angle between the thrust axis~\cite{thrust} of the $B_s$ candidate 
and that of the rest of the event, to discriminate between the two cases. 
The distribution is strongly peaked near $|\cos\theta_{\rm thr}|=1.0$ for 
$q\bar{q}$ events and is nearly flat in $\cos\theta_{\rm thr}$ for 
$B_s^{(*)}\bar{B}_s^{(*)}$ events. We require $|\cos\theta_{\rm thr}|<0.80$ 
for the $B_s\to D_s^{(*)}\pi$ final states; this eliminates about 83\% of 
the continuum background and retains 79\% of the signal events.


\section{Analysis of the $\UFS$ data}

\begin{figure}[!t]
  \includegraphics[width=0.48\textwidth]{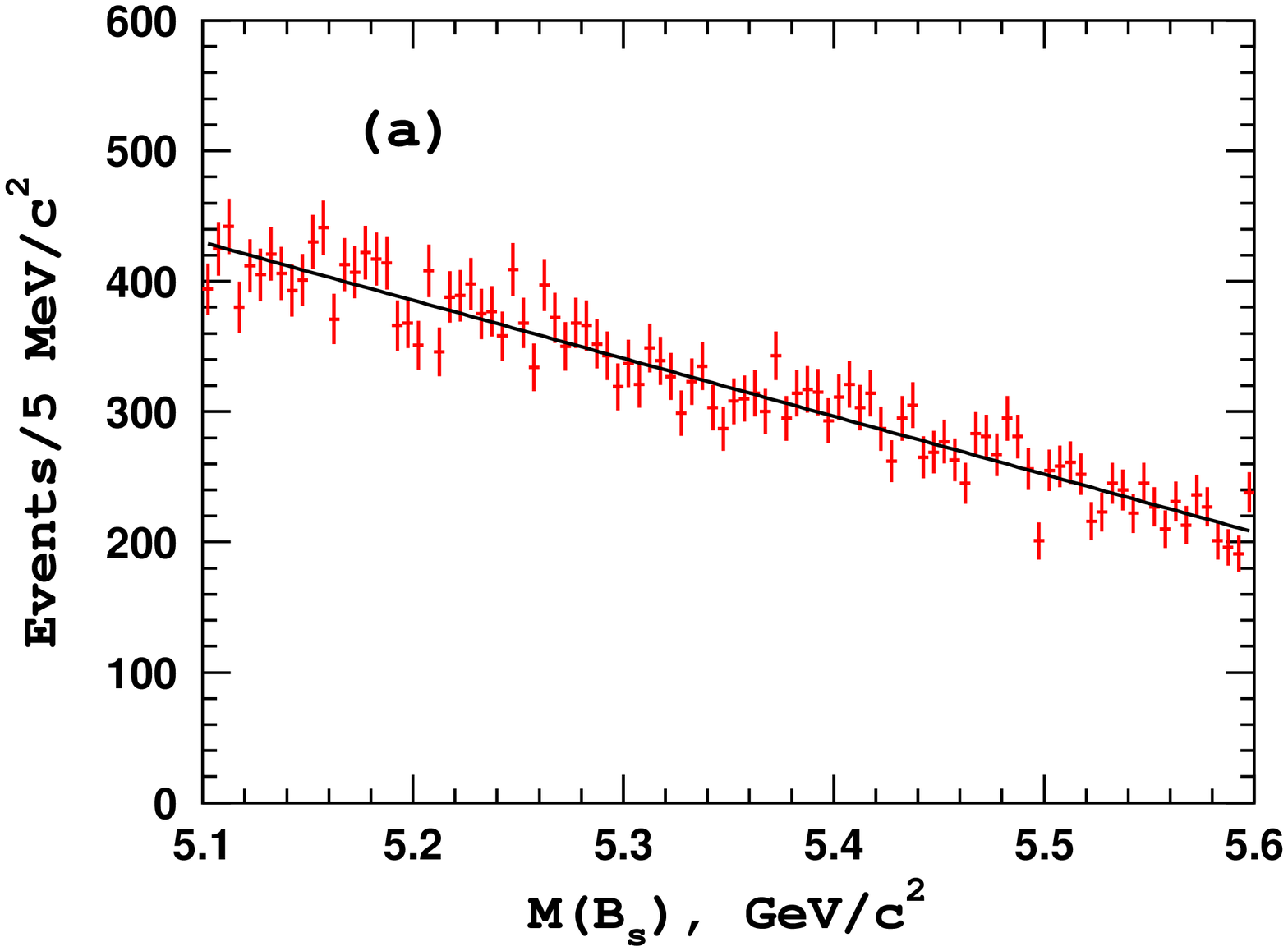} \hfill
  \includegraphics[width=0.48\textwidth]{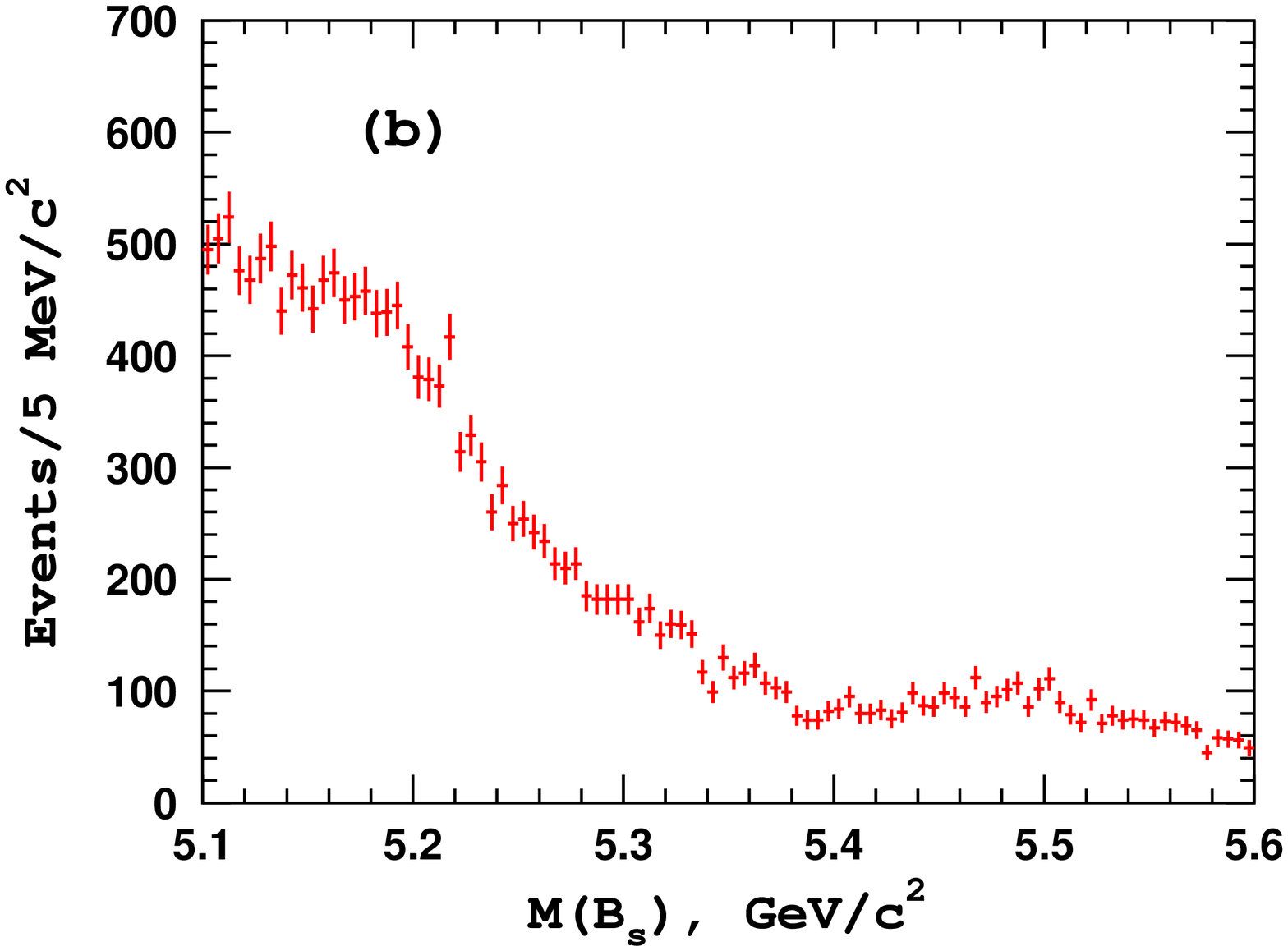} \\
  \includegraphics[width=0.48\textwidth]{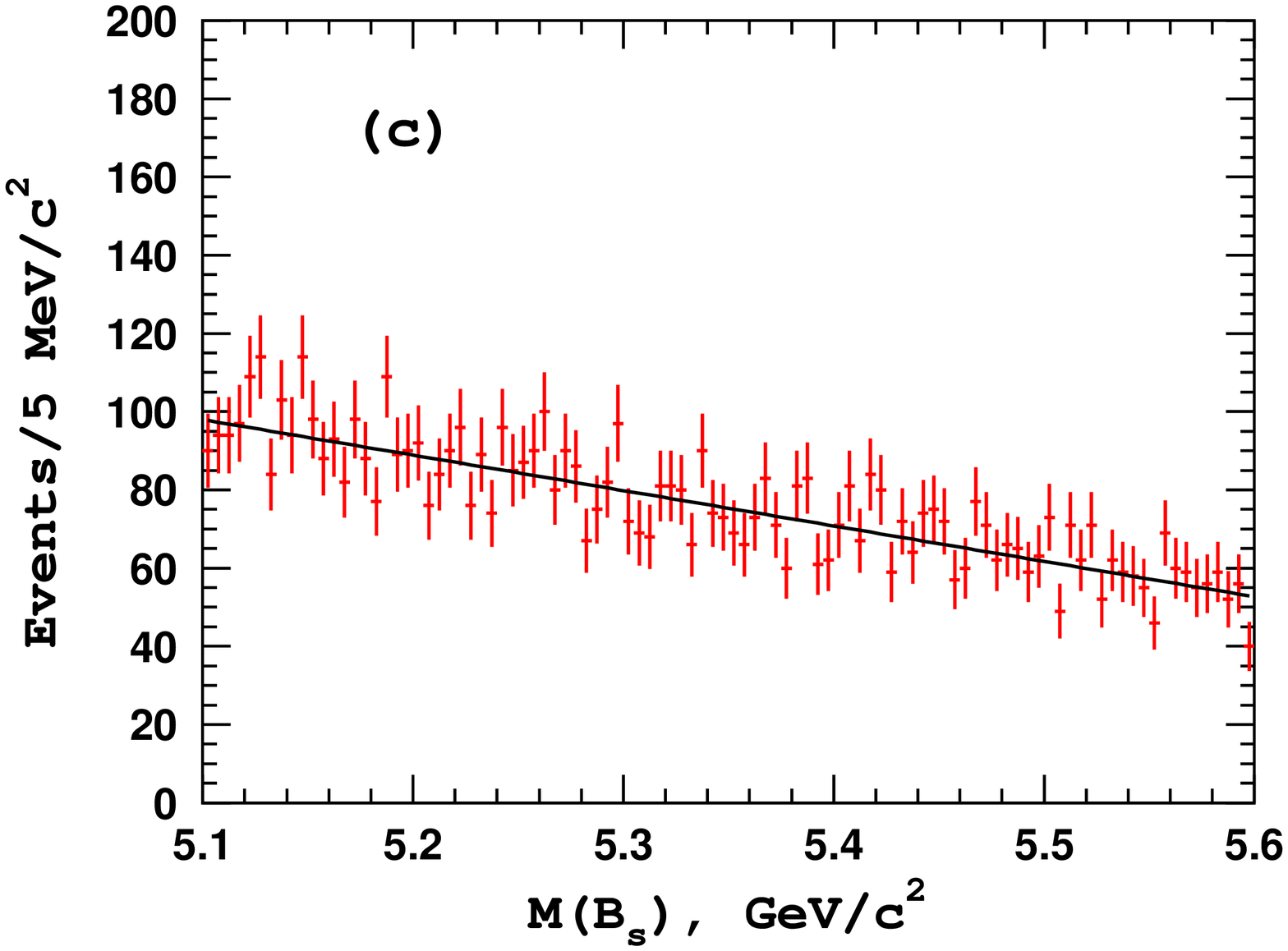} \hfill
  \includegraphics[width=0.48\textwidth]{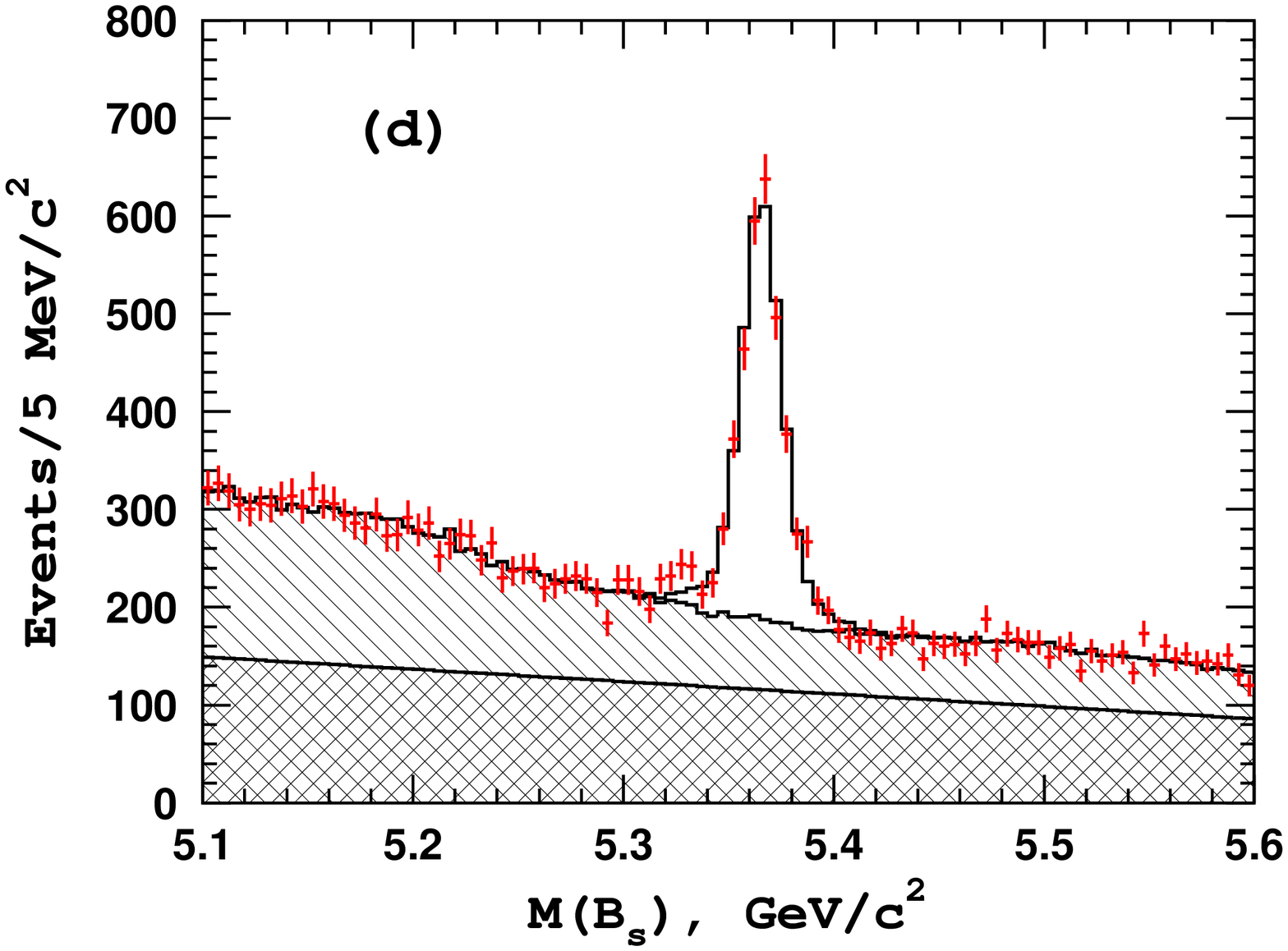} \\
  \caption{Mass distribution for the selected $B_s$ candidates
           (all modes combined) in the (a) $B_u$ and $B_d$ generic MC,
           (b) $B_s$ generic MC except for signal modes, 
           (c) continuum $\ee\to q\bar{q}$ generic MC, and 
           (d) $\UFS$ data.
The black histogram in (d) represents result of the fit with the signal 
component shown by the open histogram, $B$- and $B_s$-related background 
by the hatched histogram, and the continuum background by the 
cross-hatched histogram.}

  \label{fig:mbs}
\end{figure}

Figures~\ref{fig:mbs}(a), (b), and (c) show the combined $M(B_s)$ 
distribution for the generic $\UFS\to B^{(*)}\bar{B}^{(*)}$ MC, generic
$\UFS\to B_s^{(*)}\bar{B}_s^{(*)}$ MC (with signal modes removed), and 
continuum  $\UFS\to q\bar{q}$ ($q=u,d,s,c$) MC, respectively, with a 
requirement on the $B_s$ candidate momentum of $P(B_s)<0.95$~GeV/$c$. 

The combined $M(B_s)$ distribution for the selected $B_s$ candidates in 
data is shown in Fig.~\ref{fig:mbs}(d). To determine the $B_s$ signal 
yield, we perform a binned maximum likelihood fit of the $M(B_s)$ 
distribution to the non-coherent sum of signal and background components. 
The signal is parametrized by the sum of two Gaussian functions with a 
common mean, a ratio of widths fixed from the signal MC at 
$\sigma_2=2.1\sigma_1$, and a relative area of $N_2=0.36N_1$. 
The background component is comprised of the continuum background 
and the $B$- and $B_s$-related background. As evident from 
Figs.~\ref{fig:mbs}(a) and (c), the $B$-related and continuum backgrounds 
are featureless, so we parametrize these by linear functions. The shape 
of the $B_s$-related background, shown in Fig.~\ref{fig:mbs}(b), is fixed 
from the generic MC, while the normalization is fixed to be a fraction 
of the observed $B_s$ signal. The ratio of the number of the background 
events 
due to other $B_s$ decays to the number of events in the $B_s$ peak is 
determined to be $1.87$ for the $P(B_s)$ requirement used to select a 
combination of $B^{(*)}_s\bar{B}^{(*)}_s$ final states and $1.12$ for 
the $B_s^*\bar{B}_s^*$ final state. If the normalization is allowed to 
float while fitting the data, the fits yield $1.82\pm0.22$ and 
$1.06\pm0.13$, respectively. The result of the fit to the $M(B_s)$ 
distribution is shown in Fig.~\ref{fig:mbs}(d).
The fit yields $2283\pm 63$ signal $B_s$ decays.

\begin{figure}[!t]
  \includegraphics[width=0.48\textwidth]{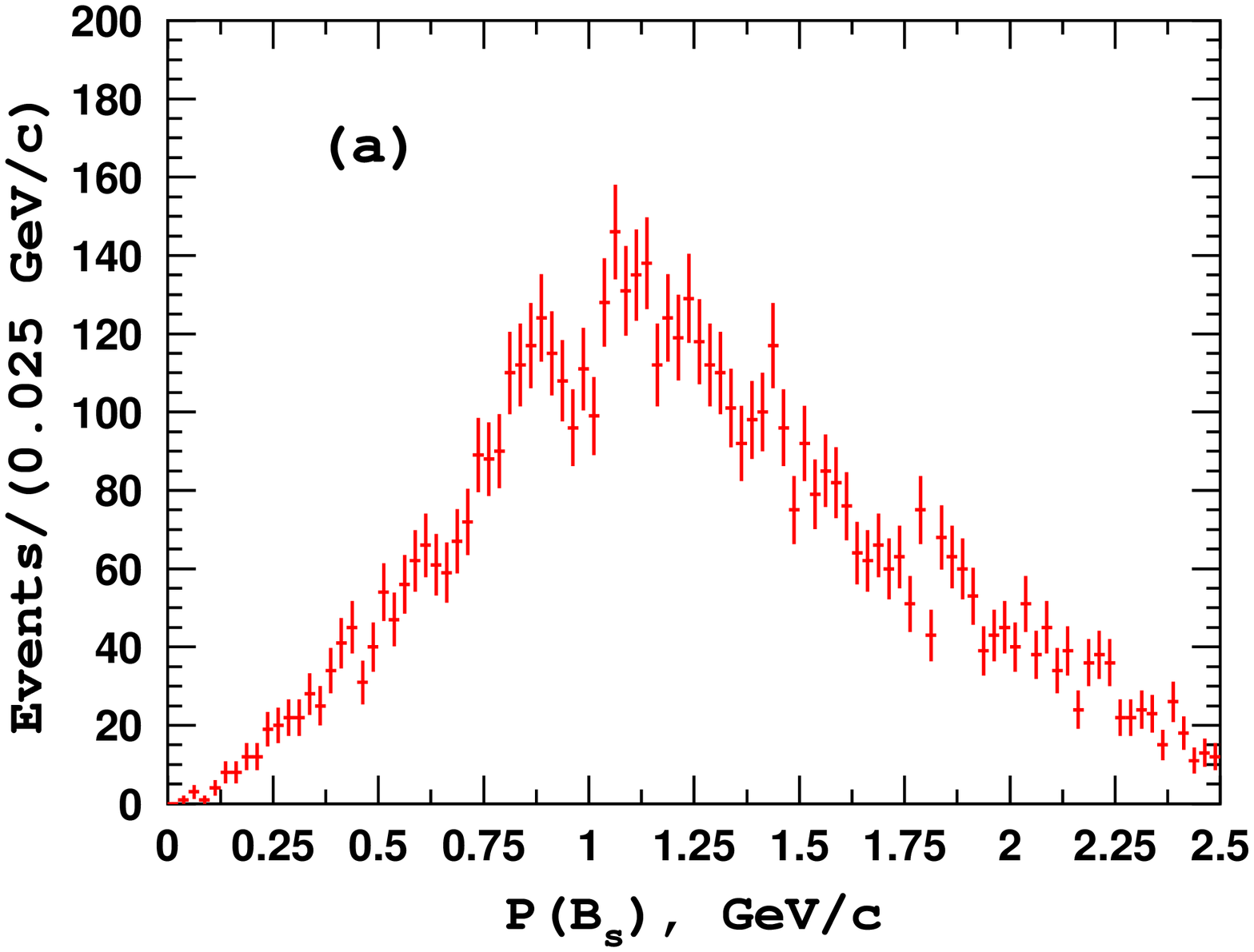} \hfill
  \includegraphics[width=0.48\textwidth]{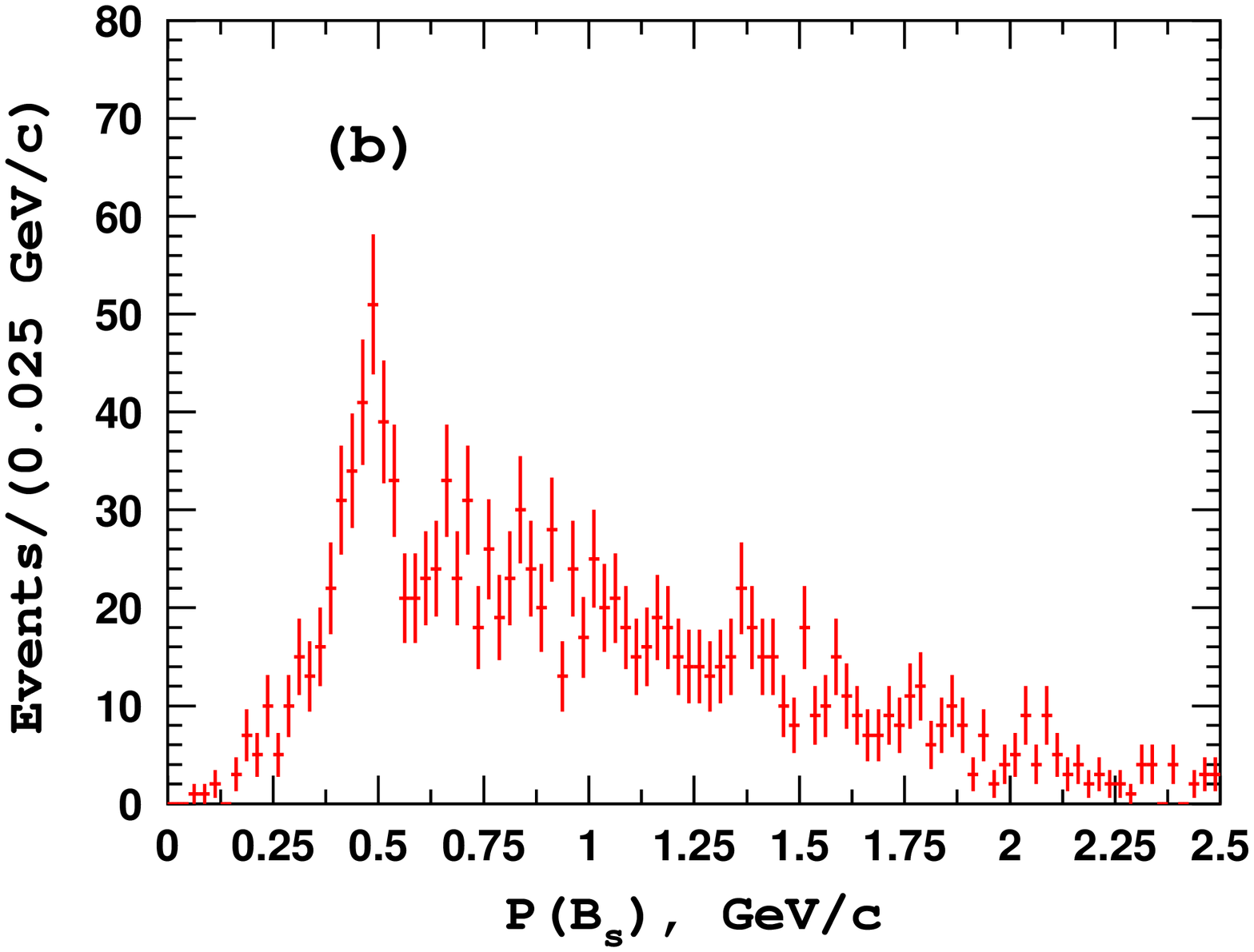} \\
  \includegraphics[width=0.48\textwidth]{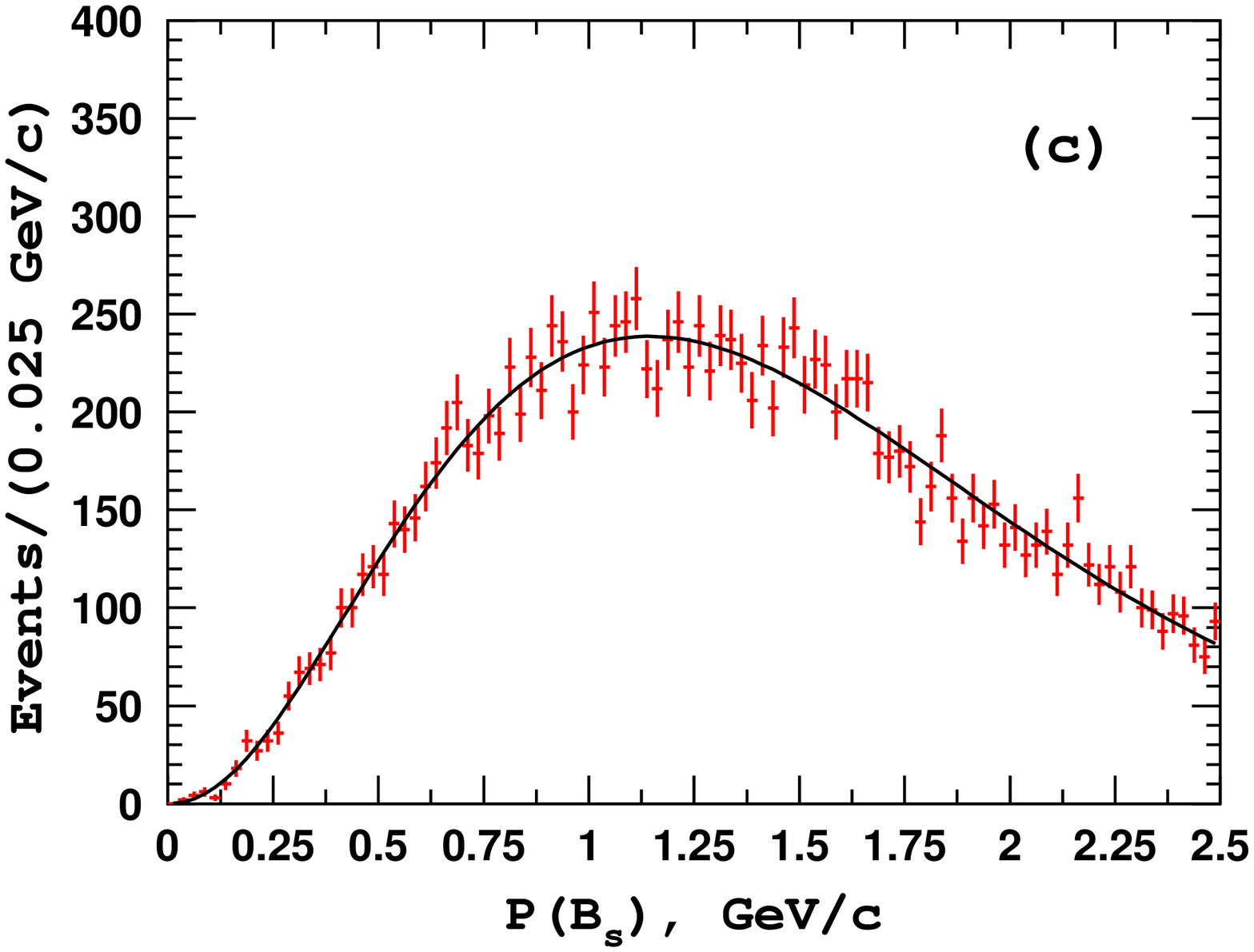} \hfill
  \includegraphics[width=0.48\textwidth]{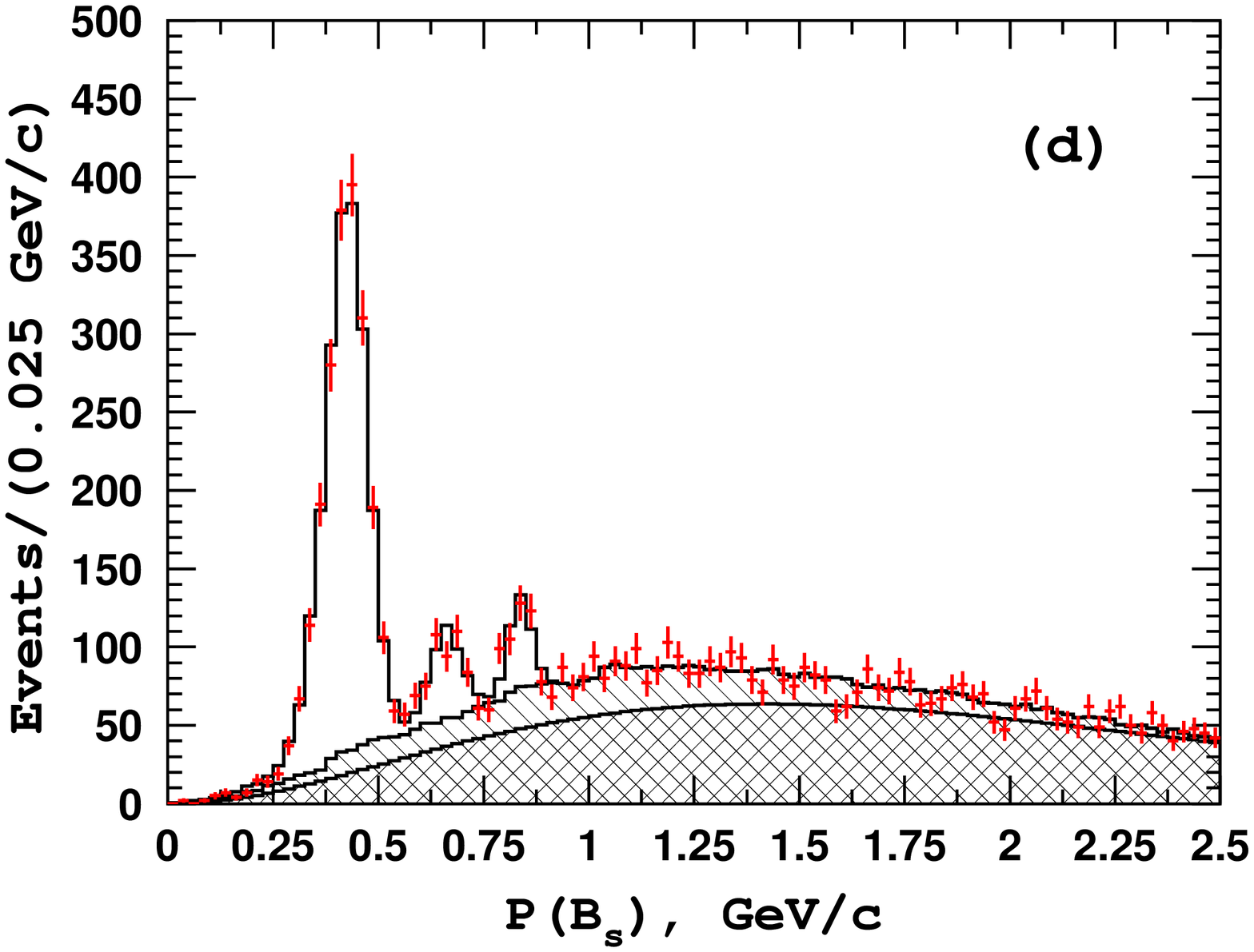} \\
  \caption{Momentum distribution for the selected $B_s$ candidates
           (all modes combined) in the (a) $B_u$ and $B_d$ generic MC,
           (b) $B_s$ generic MC with signal modes removed, 
           (c) continuum $\ee\to q\bar{q}$ generic MC, and 
           (d) $\UFS$ data. 
The black histogram in (d) represents a result of the fit with the 
signal component shown by the open histogram, $B$- and $B_s$-related 
background by the hatched histogram, and the continuum background by 
the cross-hatched histogram.}
  \label{fig:pbs}
\end{figure}

To distinguish between individual two-body $\ee\to B_s^{(*)}\bar{B}_s^{(*)}$ 
processes, we impose a requirement on the invariant mass of the $B_s$
candidate equivalent to a Gaussian $2.5\sigma$ efficiency, where $\sigma$ 
is a $B_s$ decay mode-dependent parameter. Figures~\ref{fig:pbs}(a), (b),
and (c) show the $P(B_s)$ distribution for the generic 
$\UFS\to B^{(*)}\bar{B}^{(*)}$ MC, generic $\UFS\to B_s^{(*)}\bar{B}_s^{(*)}$ 
MC (with signal modes removed), and continuum $\UFS\to q\bar{q}$ MC, 
respectively, with a $B_s$ decay mode-dependent requirement on the $M(B_s)$ 
that corresponds to a Gaussian $2.5\sigma$ efficiency. A peaking structure 
observed in Fig.~\ref{fig:pbs}(b) around $P(B_s)\sim0.5$~GeV/$c$ is due to 
misreconstructed $B_s$ candidates, such as $B^0_s\to D^-_s\pi^+$, 
$D^-_s\to K^+K^-\pi^-$ with double $\pi/K$ misidentification. Such events 
produce no peak in the $M(B_s)$ distribution but do peak in $P(B_s)$. 
The momentum distribution for the selected $B_s$ candidates in data is 
shown in Fig.~\ref{fig:pbs}(d). Three distinct peaks, corresponding to the 
$B_s\bar{B}_s$, $B_s\bar{B}_s^*+\bar{B}_sB_s^*$, and $B_s^*\bar{B}_s^*$ 
final states, are apparent.

We perform a binned maximum likelihood fit of the $P(B_s)$ distribution
to the non-coherent sum of three signal components and a background component. 
The shape of each signal component is determined from MC simulation with 
the initial state radiation (ISR) effect taken into account. The background 
component is comprised of the continuum background, the $B$-related 
background, and the $B_s$-related background. The shape of the continuum 
$P(B_s)$ background is parametrized as
\begin{equation}
B_{qq}(x) \sim x^\alpha e^{-(x/x_0)^\beta},
\end{equation}
where $x = P(B_s)$; $x_0$, $\alpha$, and $\beta$ are fit parameters.
The normalization of the continuum background component is allowed to 
float. For the $B$- and $B_s$-related background components, we use the 
corresponding MC drived histograms (see Fig.~\ref{fig:pbs}) as PDFs. 
The ratios of the $B$- and $B_s$-related backgrounds to the $B_s^{(*)}$ 
signal yield are fixed from the MC simulation.

Results of the fit to the $P(B_s)$ distribution are shown in 
Fig.~\ref{fig:pbs}(d). The fit yields $1854\pm 51$ $B_s^*\bar{B}_s^*$ 
signal events, $226\pm 27$ $B_s\bar{B}_s^*+\bar{B}_sB_s^*$ signal 
events, and $169\pm 24$ $B_s\bar{B}_s$ signal events. Assuming a uniform 
reconstruction efficiency over the relevant $B_s$ momentum range, this 
corresponds to relative fractions of
$7:$ $0.853\pm0.106(stat.)\pm0.053(syst.):$ 
$0.638\pm0.094(stat.)\pm0.033(syst.)$. 
These can be compared with the current world average results of
$7:$ $0.537\pm0.152:$ $0.199\pm0.199$~\cite{PDG} and an expectation of 
$7:4:1$ in the heavy-quark spin symmetry (HQSS) 
approximation~\cite{spin,Vol1}.

\begin{table}[!t]
  \caption{Summary of the systematic studies for the relative
$B_s^*\bar{B}_s^*:$ $B_s\bar{B}_s^*+\bar{B}_sB_s^*:$ $B_s\bar{B}_s$ 
yields.}
  \medskip
  \label{tab:syst1}
\centering
  \begin{tabular}{l|ccc|c|cc} \hline \hline
 Source & \multicolumn{3}{c|}{Signal yield, events} & {Ratio} & \multicolumn{2}{c}{Uncertainty}\\
        & $B_s^*\bar{B}_s^*$ &  $B_s\bar{B}_s^*$  &  $B_s\bar{B}_s$ &  & $B_s\bar{B}_s^*$ &  $B_s\bar{B}_s$\\
  \hline
 $B$\&$B_s$ background  & & & &  \\
 ~~~floating                   & 1865 & 219 & 168 & $7:0.822:0.637$ & \\
 ~~~$\times 1.50$              & 1844 & 227 & 164 & $7:0.862:0.623$ & \\
 ~~~$\times 0.75$              & 1863 & 221 & 172 & $7:0.830:0.646$ & \\
& & & & & $^{+0.009}_{-0.021}$ & $^{+0.008}_{-0.015}$ \\ 
 \hline
 $M(B_s)$ signal region & & & & \\
 ~~~$\pm 2\sigma$              & 1780 & 212 & 162 & $7:0.834:0.637$ & \\
 ~~~$\pm 3\sigma$              & 1897 & 235 & 174 & $7:0.867:0.642$ & \\
& & & & & $^{+0.014}_{-0.019}$ & $^{+0.004}_{-0.001}$ \\
 \hline
 $P(B_s)$ range         & & & & \\
 ~~~$<2.00$~GeV/$c$            & 1864 & 226 & 165 & $7:0.851:0.626$ & \\
 ~~~$<2.25$~GeV/$c$            & 1857 & 225 & 167 & $7:0.851:0.636$ & \\
 ~~~$<2.75$~GeV/$c$            & 1859 & 222 & 165 & $7:0.838:0.628$ & \\
 ~~~$<3.00$~GeV/$c$            & 1871 & 231 & 173 & $7:0.871:0.647$ & \\
& & & & & $^{+0.018}_{-0.015}$ & $^{+0.005}_{-0.016}$ \\
 \hline
 Momentum resolution    & & & & \\
 ~~~$B_s^*\bar{B}_s^*:$ $-10$\% & 1842 & 213 & 162 & $7:0.811:0.622$ & \\ 
 ~~~$B_s^*\bar{B}_s^*:$ $+10$\% & 1865 & 239 & 177 & $7:0.900:0.671$ & \\ 
 ~~~$B_s\bar{B}_s^*:$   $-10$\% & 1855 & 226 & 169 & $7:0.855:0.644$ & \\
 ~~~$B_s\bar{B}_s^*:$   $+10$\% & 1856 & 218 & 162 & $7:0.824:0.617$ & \\
 ~~~$B_s\bar{B}_s:$     $-10$\% & 1854 & 227 & 171 & $7:0.860:0.652$ & \\
 ~~~$B_s\bar{B}_s:$     $+10$\% & 1854 & 224 & 166 & $7:0.848:0.633$ & \\
& & & & & $^{+0.047}_{-0.042}$ & $^{+0.029}_{-0.025}$ \\
 \hline \hline
 Nominal fit & $1854\pm51$ &  $226\pm27$  &  $169\pm24$ & $7:$ $0.853\pm0.106:$ $0.638\pm0.094$ &
$^{+0.053}_{-0.053}$ & $^{+0.030}_{-0.033}$ \\
  \hline  \hline
  \end{tabular}
\end{table}

The dominant sources of the systematic uncertainties for the 
relative fractions of the two-body signals are:
\begin{itemize}
  \item the fraction of the $B$- and $B_s$-related background
estimating by repeating the fit to the $B_s$ momentum distribution with 
the normalization of this background allowed to float;
  \item the $M(B_s)$ signal region, estimated by repeating the fit to the 
data with the $M(B_s)$ signal region set to $\pm3\sigma$ and $\pm 2\sigma$
around the $B_s$ nominal mass;
  \item the momentum distribution fitting range, estimated by varying the 
upper boundary of the momentum range from 2.0 to $3.0$~GeV/$c$ with a 
0.25~GeV/$c$ step;
  \item the width of the momentum resolution function, estimated by 
varying the width of the $P(B_s)$ resolution within $\pm10$\% of the 
nominal value and repeating the fit to the data.
\end{itemize}
These uncertainties are summarized in Table~\ref{tab:syst1}.
The overall systematic uncertainty is estimated to be $\pm0.053$
for the $B_s\bar{B}_s^*+\bar{B}_sB_s^*$ fraction and $\pm0.033$
for the $B_s\bar{B}_s$ fraction.


\subsection{$B_s$ reconstruction efficiency}

\begin{figure}[!t]
  \includegraphics[width=0.49\textwidth]{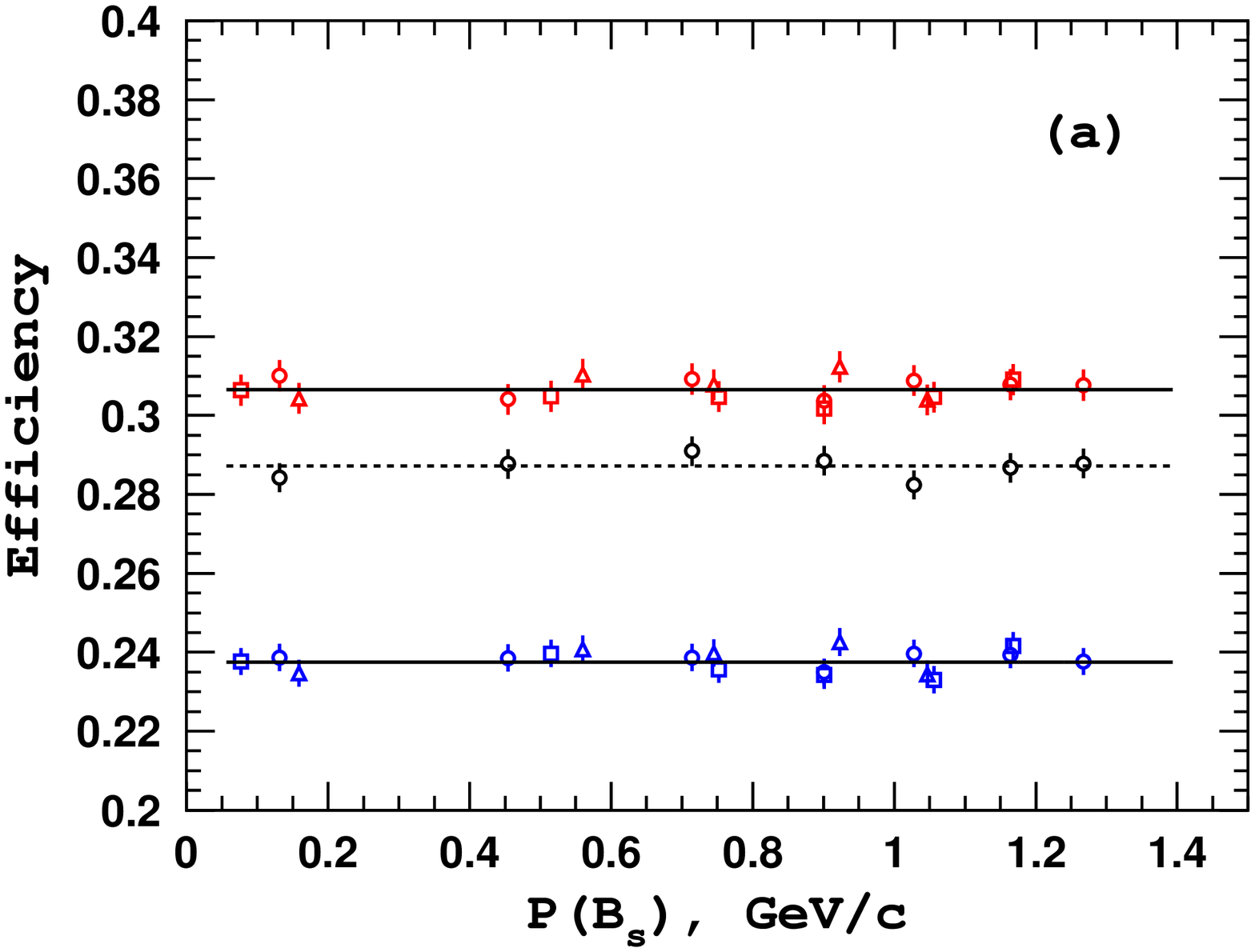} \hfill
  \includegraphics[width=0.49\textwidth]{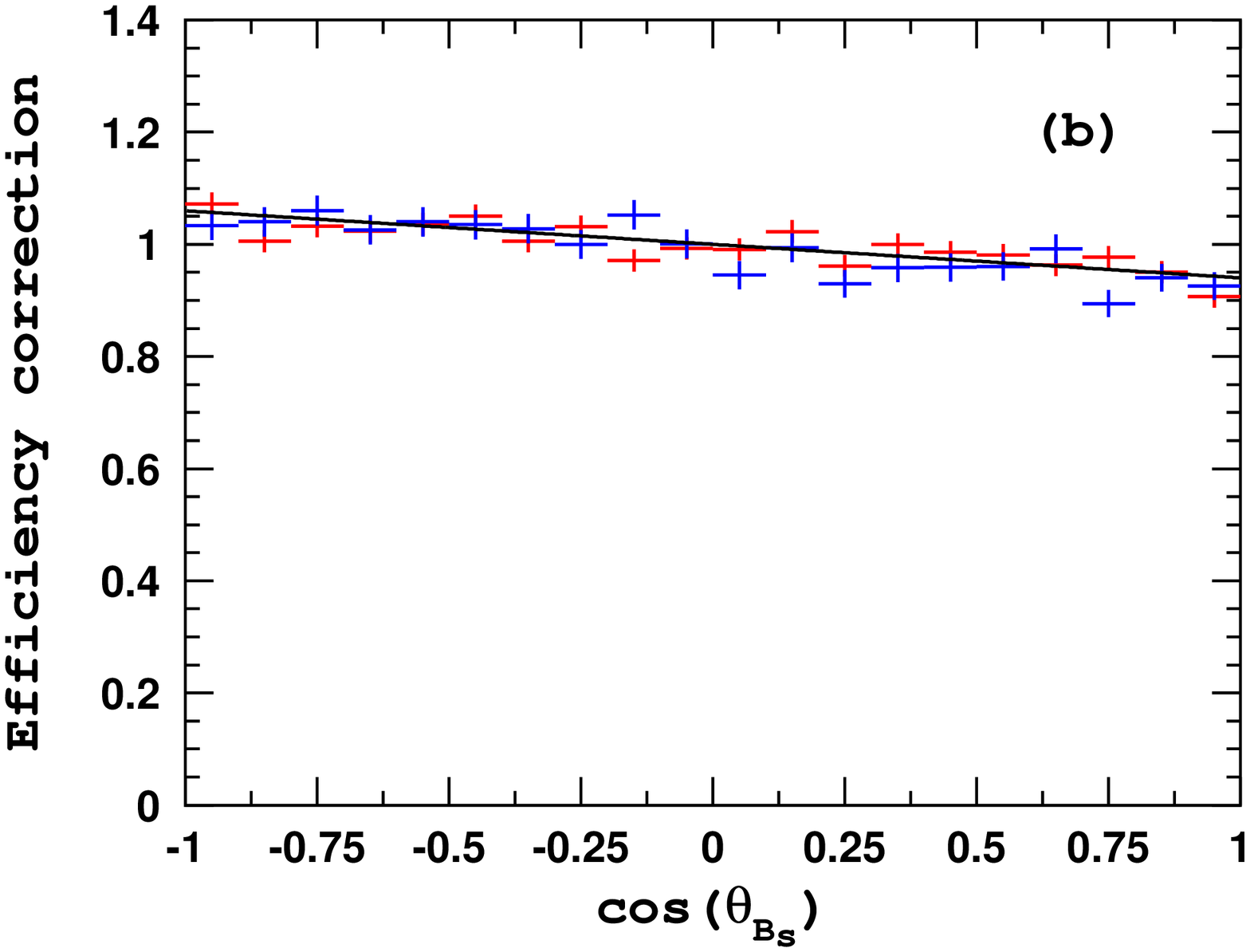} 
  \caption{$B_s$ reconstruction efficiency (no intermediate branching 
    fractions included). (a) Momentum dependence of the $B_s$ 
    reconstruction efficiency for the $B_s\to D_s[K^+K^-\pi]\pi$ 
    decay mode with no $\cos(\theta_{\rm thr})$ cut (red points), 
    with the $|\cos(\theta_{\rm thr})|<0.8$ cut applied (blue points), and
    for the  $B_s\to J/\psi[\ell^+\ell^-]K^+K^-$ decay mode (black points).
    (b) Correction for the $B_s$ reconstruction efficiency as a function 
    of the $B_s$ polar angle in the c.m.\ frame. Red points are for the 
    $B_s\to D_s[K^+K^-\pi]\pi$ decay mode, blue points are for the 
    $B_s\to J/\psi[\mu^+\mu^-]K^+K^-$ decay mode. The solid line represents
    the result of the fit to a linear function.}
  \label{fig:eff}
\end{figure}

To account for the possible dependence of the $B_s$ reconstruction 
efficiency on the c.m. energy ($P(B_s)$ momentum), we generate 20K
$\ee\to B_s^{(*)}\bar{B}_s^{(*)}$ signal MC events at seven $E_{\rm cm}$ 
points. Applying the same reconstruction and analysis algorithm, we 
determine the $B_s$ signal yield. The results are summarized in 
Fig.~\ref{fig:eff}. No significant variations in the reconstruction 
efficiency are observed within the relevant $B_s$ momentum range, 
including the case where the $\cos(\theta_{\rm thr})$ requirement is 
applied. 


\subsection{Angular analysis}

The $\cos(\theta_{B^*_s})$ distribution, where $\theta_{B^*_s}$ is the angle 
between the $B^*_s$ momentum and 
the $z$ axis in the 
c.m.\ frame, provides information on the relative fractions of the $S=0$ 
and $S=2$ states, with $S$ being the total spin of the $B^*_s\bar{B}^*_s$ 
pair, produced in the $\ee\to B^*_s\bar{B}^*_s$ process. The angular 
distribution of the $S=0$ component is proportional to 
$1-\cos^2(\theta_{B^*_s})$ while that for the $S=2$ component to 
$(7-\cos^2(\theta_{B^*_s}))/10$. The differential cross section then reads 
as
\begin{equation} 
 \frac{d\sigma}{d\cos(\theta_{B^*_s})} \sim {\cal{A}}^2_0 + {\cal{A}}^2_2,
\end{equation}
where ${\cal{A}}^2_0=a^2_0(1-\cos^2\theta_{B^*_s})$ and 
${\cal{A}}^2_2=a^2_2(7-\cos^2\theta_{B^*_s})/10$ are the squared amplitudes 
for the 
$B^*_s\bar{B}^*_s$ production in a $P$ wave with the total spin of 
$S=0$ and $S=2$, respectively. In the heavy quark spin symmetry, the 
ratio $a_0^2:a_2^2$ is expected to be 1:20. However, 
the proximity of the $B^*_s\bar{B}^*_s$ production threshold might 
distort this ratio significantly~\cite{Vol2}.

For the analysis of the $B^*_s$ polar angular distribution in data, we 
select $B^*_s$ candidates by applying a requirement on the $B_s$ momentum 
of $0.25$~GeV/$c<P(B_s)<0.55$~GeV/$c$ and then determine the $B_s$ yield 
in $\cos(\theta_{B_s})$ bins. (In fact, we measure the polar angle of the 
$B_s$ meson, not $B^*_s$. The associated absolute uncertainty in 
$\cos\theta_{B^*_s}$ is below 0.01, which is much smaller than the bin 
width.) We perform a binned maximum likelihood fit to the $M(B_s)$ 
distribution for each $\cos\theta_{B_s}$ bin. The $B_s$ yield as a function 
of $\cos\theta_{B_s}$ is fit to the following function:
\begin{equation}
 \frac{d\sigma}{d\cos(\theta_{B_s})} \sim r(1-\cos^2\theta_{B_s}) + 
                   (1-r)\frac{7-\cos^2\theta_{B_s}}{10},
\end{equation}
where $r=a_0^2/(a_2^2+a_0^2)$. 
We also apply the efficiency corrections described earlier.

\begin{figure}[!t]
  \includegraphics[width=0.325\textwidth]{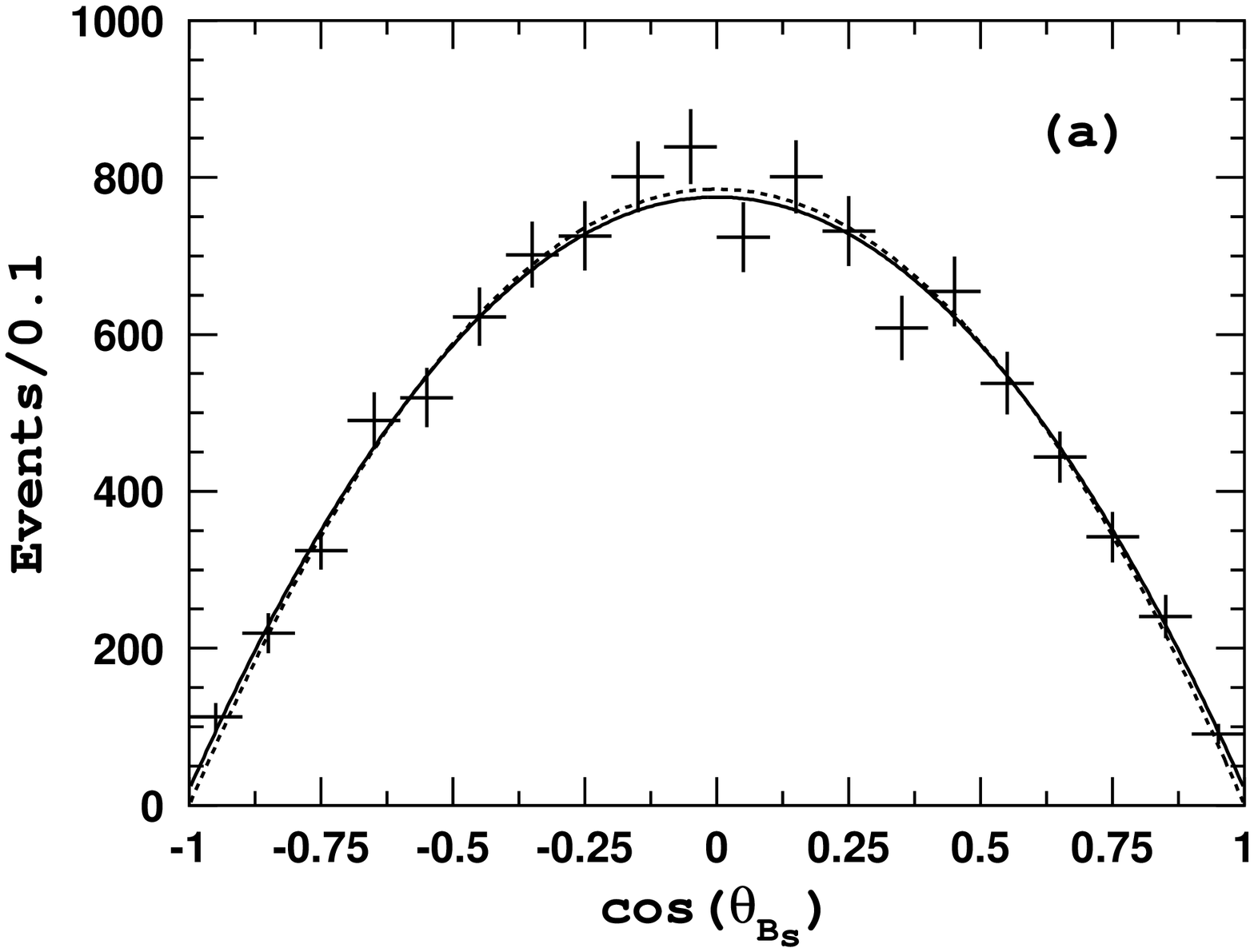} \hfill
  \includegraphics[width=0.325\textwidth]{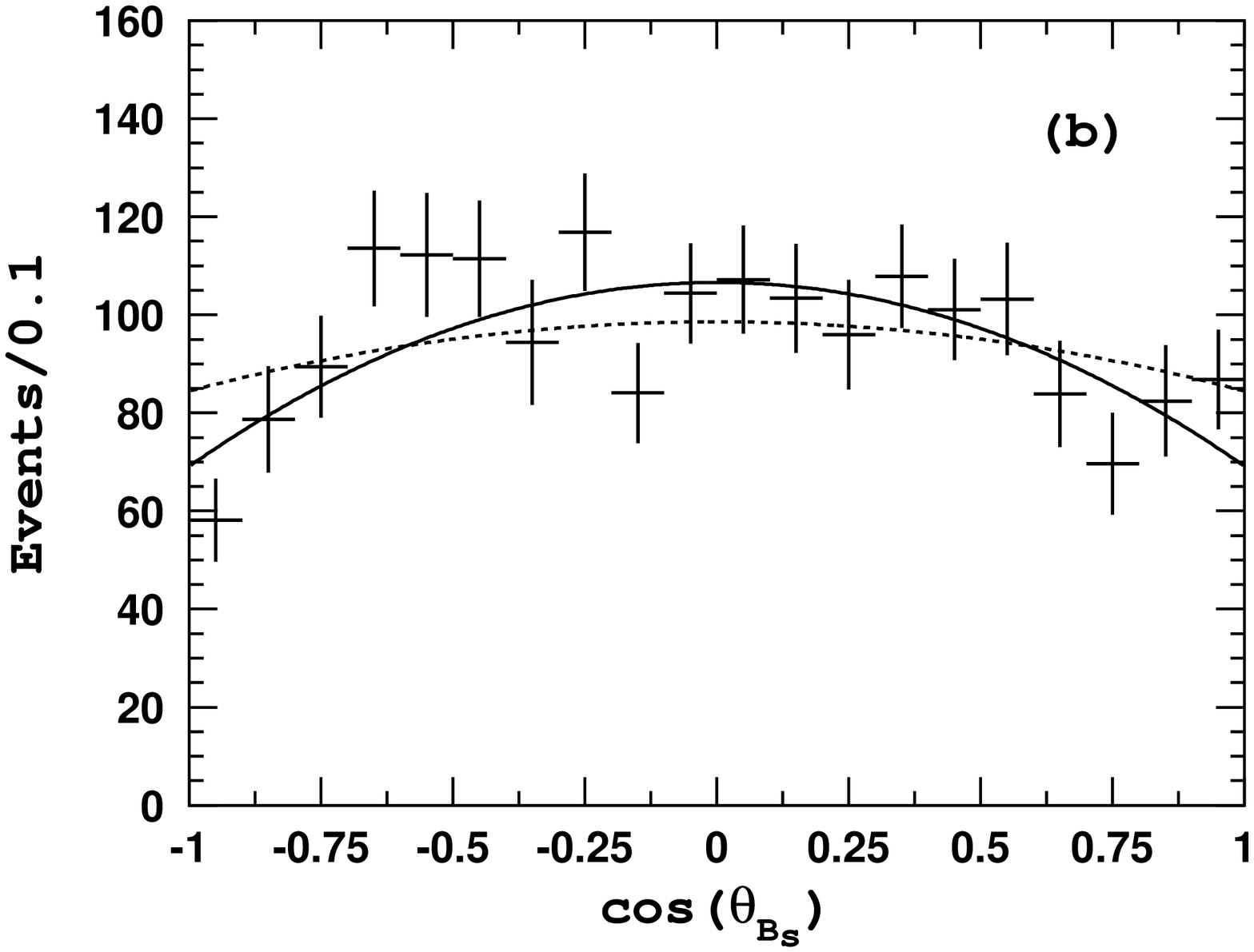} \hfill
  \includegraphics[width=0.325\textwidth]{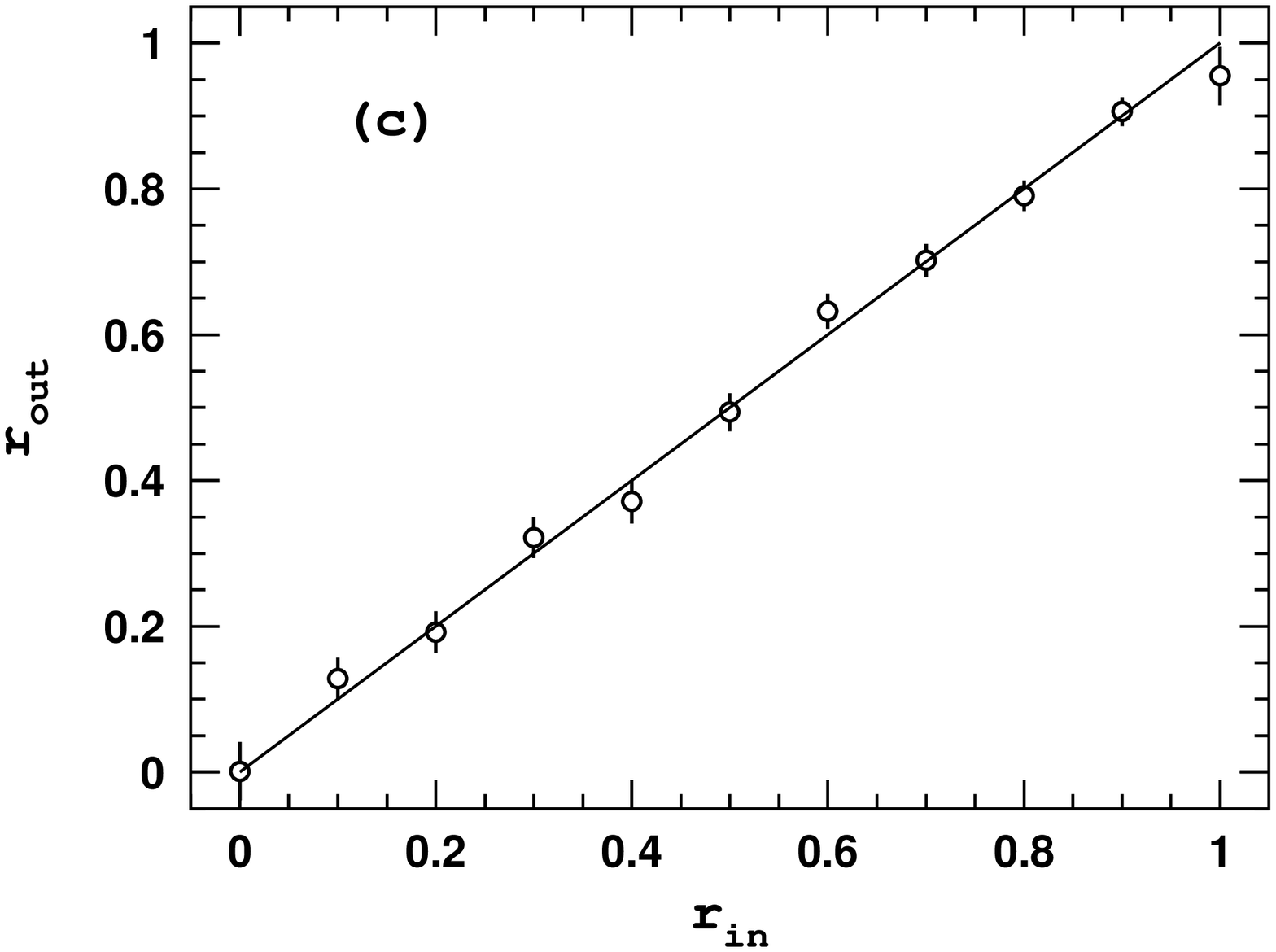} 
  \caption{$\cos(\theta_{B_s})$ distribution for the 
  $\ee\to B^*_s\bar{B}^*_s$ events.
  (a) Test with generic MC events. The solid line -- a fit with combinations 
  of the $S=0$ and $S=2$ components; the dashed line -- a fit with the 
  $S=0$ component only.
  (b) $\UFS$ data. Th solid line  -- a fit with combinations of
  the $S=0$ and $S=2$ components; the dashed line -- fit with the $S=2$ 
  component only.
  (c) measured $r$ value versus the input one as determined with signal MC;
  the solid line shows the exact proportionality.}
  \label{fig:theta}
\end{figure}

As a cross-check of the analysis procedure, we apply it
to the generic MC events. Results of this analysis are shown in 
Fig.~\ref{fig:theta}(a). The fit result of $r=0.952\pm0.029$ is 
consistent with a pure $S=0$ component. This agrees with the MC 
input, where the fraction of the $S=2$ component is (wrongly) set 
to zero. 

Results of the same analysis applied to the data are shown in 
Fig.~\ref{fig:theta}(b). The fit yields a fraction of the $S=0$
component of $r=0.175\pm0.057^{+0.022}_{-0.018}$. We also fit the data 
with a pure $S=2$ form, the results are also shown in 
Fig.~\ref{fig:theta}(b). The statistical significance of the $S=0$ 
component, determined as 
$\sqrt{-2(\ln{\cal{L}}_{S=2}-\ln{\cal{L}}_{\rm mix})}$ is 3.1 
standard deviations (statistical only). 

The dominant sources of the systematic uncertainties for the 
angular analysis are 
\begin{itemize}
  \item correction for the reconstruction efficiency -- 
$^{+0.004}_{-0.000}$: to estimate this uncertainty, we vary the slope 
of the correction function within its statistical uncertainty;
  \item binning -- $\pm0.010$: to estimate this uncertainty, we repeat 
the fit with bin widths of 0.040, 0.050, 0.080, 0.125, and 0.200, then 
take the largest positive and negative deviations as the estimation of 
the systematic uncertainty;
  \item determination of the $B_s$ signal yield -- $^{+0.015}_{-0.008}$:
here, we vary the fraction of the $B_s$ related component within $\pm25$\%
and fraction of the second Gaussian in the signal PDF within $\pm10$\%
(the typical variation of these quantities for various $B_s$ decay chains) 
and repeat the fit to the angular distribution;
  \item momentum cuts to select the $B_s^*\bar{B}_s^*$ signal --
$\pm 0.012$: here, we vary the lower and the higher boundary of the
momentum range by $\pm0.05$~GeV/$c$ and repeat the fit to the
angular distribution.
\end{itemize}

We also check for a possible systematic shift in the determination 
of the $r$ value (linearity check) using signal MC events generated
with various inputs for the $S=0$ fraction. The results of this study
are shown in Fig.~\ref{fig:theta}(c). 

The overall systematic uncertainty is calculated as the quadratic 
sum of all contributions and is $^{+0.022}_{-0.018}$. This reduces the 
significance of the $S=0$ component to $2.6\sigma$.


\section{Analysis of the Energy Scan Data}

For this analysis, we use 19 energy points above the $B_sB_s$ production 
threshold with about one inverse femtobarn of integrated luminosity 
accumulated at each point. We also split the 121.4~fb$^{-1}$ of data 
taken near the $\UFS$ peak into three samples with close $E_{\rm cm}$ 
values according to the KEKB data; see Table~\ref{tab:e-scan}.

\begin{table}[!t]
  \caption{Summary of the energy scan results.}
  \medskip
  \label{tab:e-scan}
\centering
  \begin{tabular}{rcc|ccc|ccc} \hline \hline
 \# & ~Energy~ &  Lumi. & 
\multicolumn{3}{c|}{Total $B_s^{(*)}\bar{B}_s^{(*)}$} &
\multicolumn{3}{c}{Only $B_s^*\bar{B}^*_s$} \\
& &   & $P(B_s)$ & $B_s$ Yield & $\sigma_{\rm vis}$ & 
        $P(B_s)$ & $B_s$ Yield & $\sigma_{\rm vis}$   \\
& (GeV) & (fb$^{-1}$) & (GeV/$c$) & (Events)  & (pb) &
                     (GeV/$c$) & (Events) &  (pb)
\\ \hline
 1  & 10.7711 &  0.955 & $<0.605$ & $  3.0\pm 2.3$ & $ 9.8\pm 7.5\pm 3.2$ &    $-$   &     $-$      &     $-$  \\
 2  & 10.8205 &  1.697 & $<0.793$ & $  4.8\pm 4.1$ & $ 8.8\pm 7.5\pm 2.9$ &    $-$   &     $-$      &     $-$  \\
 3  & 10.8497 &  0.989 & $<0.888$ & $ 14.3\pm 6.2$ & $45.0\pm19.5\pm 8.7$ & $<0.461$ & $12.3\pm3.3$ & $38.7\pm10.4\pm7.2$ \\
 4  & 10.8589 &  0.988 & $<0.916$ & $ 26.8\pm 6.3$ & $84.4\pm19.9\pm15.7$ & $<0.520$ & $15.8\pm3.4$ & $49.8\pm10.7\pm9.3$ \\
 5  & 10.8695 &  0.978 & $<0.947$ & $ 28.6\pm 6.2$ & $91.0\pm19.7\pm17.2$ & $<0.578$ & $20.6\pm3.9$ & $65.6\pm12.4\pm12.2$ \\
 6  & 10.8785 &  0.978 & $<0.973$ & $ 13.5\pm 5.4$ & $43.0\pm17.2\pm 8.3$ & $<0.622$ & $12.3\pm3.9$ & $39.2\pm12.4\pm7.3$ \\
 7  & 10.8836 &  1.848 & $<0.987$ & $ 24.5\pm 7.1$ & $41.3\pm12.0\pm 7.7$ & $<0.644$ & $20.5\pm5.8$ & $34.5\pm 9.8\pm6.4$ \\
 8  & 10.8889 &  0.990 & $<1.003$ & $ 10.1\pm 5.1$ & $31.8\pm16.0\pm 6.0$ & $<0.668$ & $ 4.3\pm2.8$ & $13.5\pm 8.8\pm4.5$ \\
 9  & 10.8985 &  0.983 & $<1.029$ & $ 11.2\pm 4.7$ & $35.5\pm14.9\pm 6.6$ & $<0.708$ & $ 3.3\pm2.8$ & $10.5\pm 8.9\pm3.5$ \\
10  & 10.9011 &  1.425 & $<1.036$ & $ 13.7\pm 4.9$ & $30.0\pm10.7\pm 5.8$ & $<0.718$ & $ 9.8\pm4.0$ & $21.4\pm 8.7\pm5.3$ \\
11  & 10.9077 &  0.980 & $<1.053$ & $ -2.8\pm 3.8$ & $-8.9\pm12.1\pm 4.1$ & $<0.744$ & $-1.1\pm3.5$ & $-3.5\pm11.1\pm2.6$ \\
12  & 10.9275 &  1.149 & $<1.105$ & $  5.6\pm 4.8$ & $12.1\pm13.0\pm 4.3$ & $<0.815$ & $ 4.4\pm3.4$ & $11.9\pm 9.2\pm4.2$ \\
13  & 10.9575 &  0.969 & $<1.178$ & $ -0.2\pm 3.6$ & $-0.6\pm11.6\pm 2.3$ & $<0.912$ & $ 2.3\pm3.3$ & $ 7.4\pm10.1\pm3.4$ \\
14  & 10.9775 &  0.999 & $<1.224$ & $  2.9\pm 4.7$ & $ 9.0\pm14.6\pm 3.3$ & $<0.971$ & $ 2.8\pm3.2$ & $ 8.7\pm10.0\pm3.2$ \\
15  & 10.9919 &  0.985 & $<1.258$ & $ -4.5\pm 3.3$ &$-14.2\pm10.4\pm 4.1$ & $<1.012$ & $-1.0\pm2.6$ & $-3.1\pm 8.2\pm2.5$ \\
16  & 11.0068 &  0.976 & $<1.290$ & $ -2.9\pm 4.2$ & $-9.3\pm13.4\pm 3.8$ & $<1.052$ & $-3.5\pm2.7$ & $-11.2\pm8.6\pm4.6$ \\
17  & 11.0164 &  0.771 & $<1.311$ & $ 10.4\pm 6.1$ & $42.0\pm24.6\pm 7.9$ & $<1.077$ & $ 7.7\pm4.4$ & $31.1\pm17.8\pm5.8$ \\
18  & 11.0175 &  0.859 & $<1.314$ & $  8.2\pm 5.2$ & $29.7\pm18.8\pm 5.7$ & $<1.080$ & $ 1.4\pm3.4$ & $ 5.1\pm12.3\pm3.4$ \\
19  & 11.0220 &  0.982 & $<1.323$ & $  0.8\pm 4.2$ & $ 2.5\pm13.3\pm 3.7$ & $<1.091$ & $ 0.4\pm3.9$ & $ 1.3\pm12.4\pm4.3$ \\
20  & 10.8686 & 22.938 & $<0.945$ & $457.5\pm29.0$ & $62.1\pm3.9\pm11.5$  & $<0.573$ &  $378\pm42$  & $51.3\pm5.7\pm9.5$ \\
21  & 10.8633 & 47.647 & $<0.930$ & $817.7\pm32.3$ & $53.3\pm2.1\pm 9.9$  & $<0.545$ &  $732\pm50$  & $47.8\pm3.3\pm8.9$ \\
22  & 10.8667 & 50.475 & $<0.940$ & $999.0\pm33.0$ & $61.6\pm2.0\pm11.5$  & $<0.563$ &  $820\pm53$  & $50.6\pm3.3\pm9.4$ \\
  \hline  \hline
  \end{tabular}
\end{table}

\begin{figure}[!t]
  \includegraphics[width=0.24\textwidth]{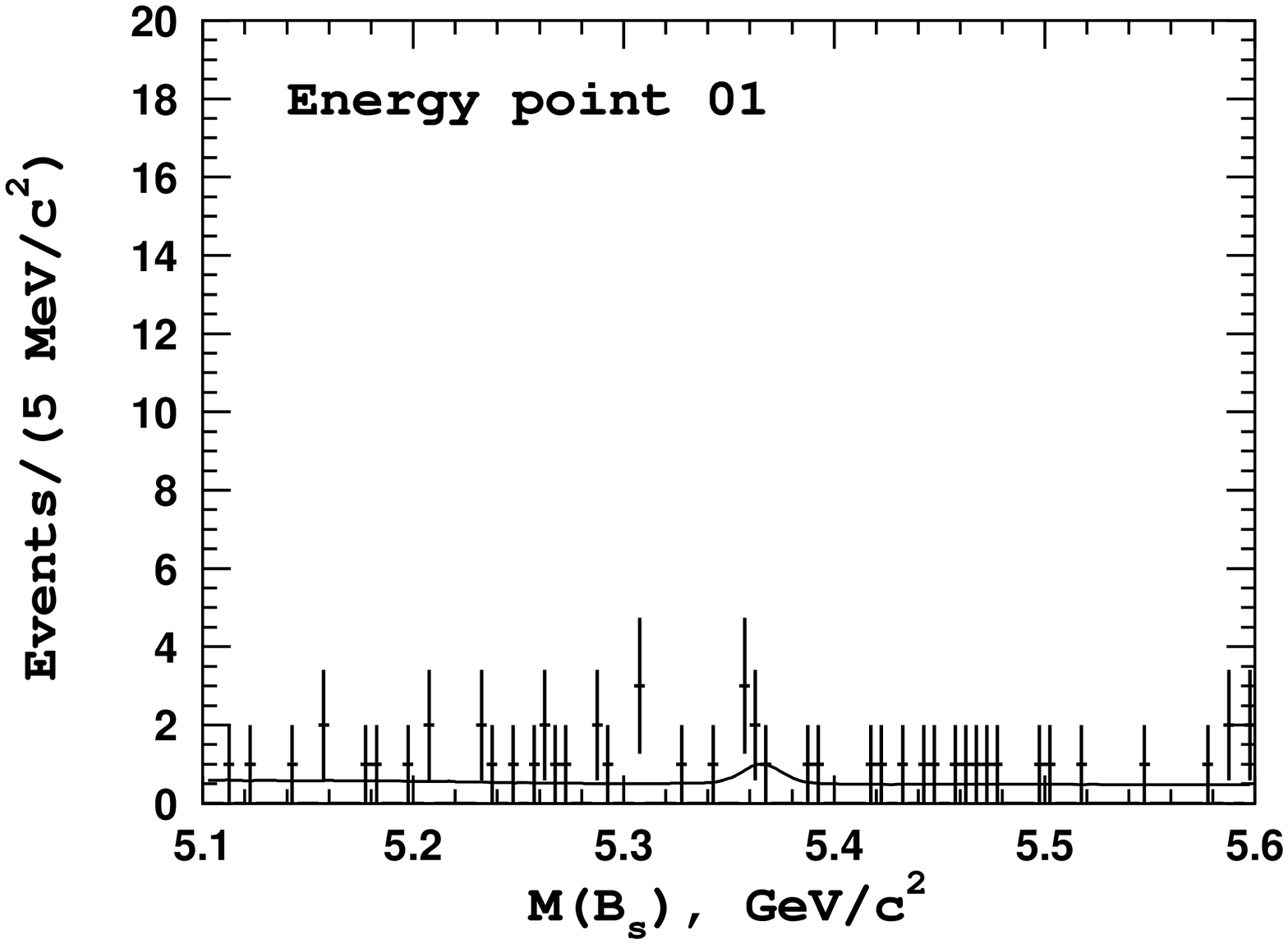} \hfill
  \includegraphics[width=0.24\textwidth]{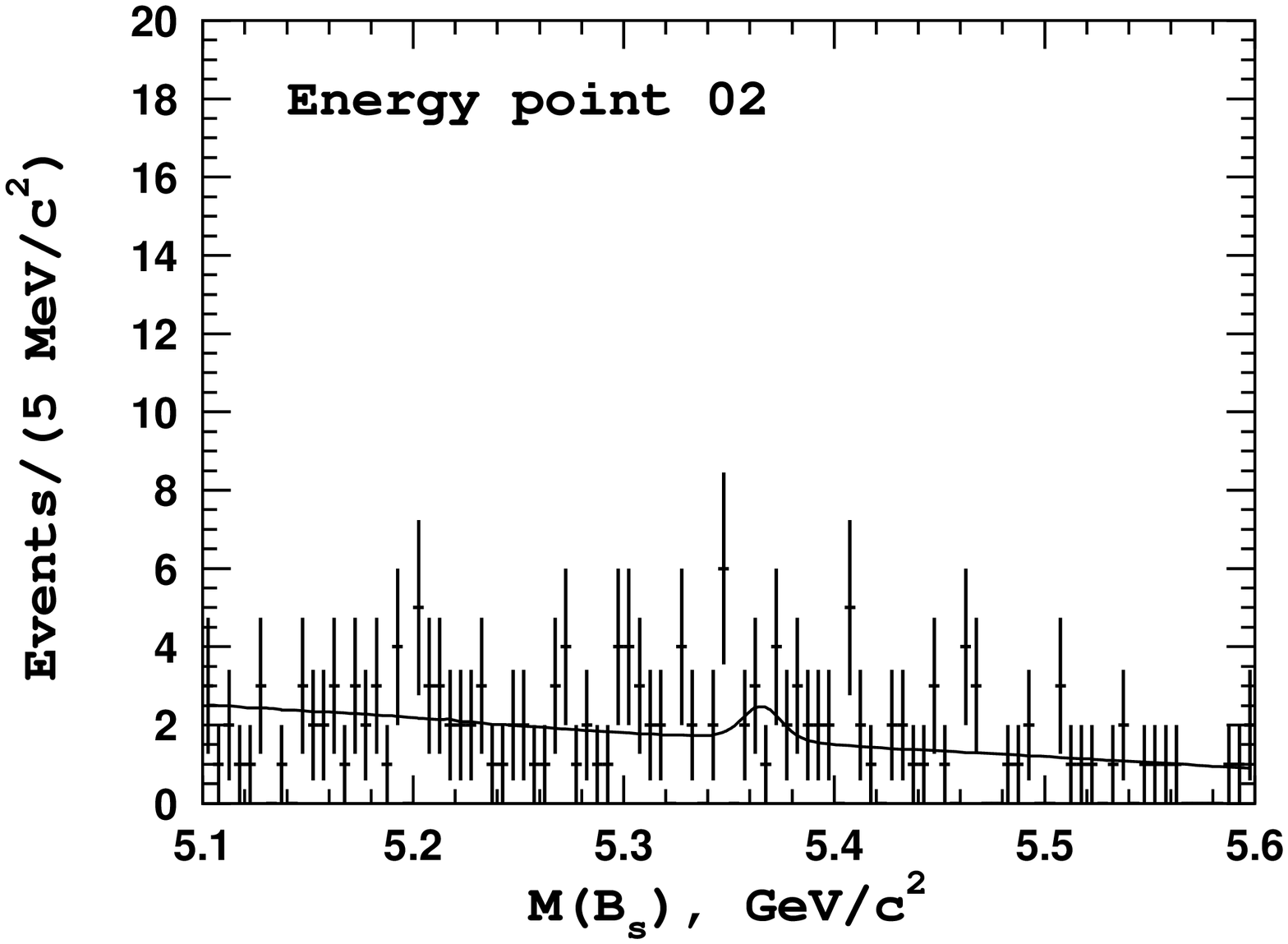} \hfill
  \includegraphics[width=0.24\textwidth]{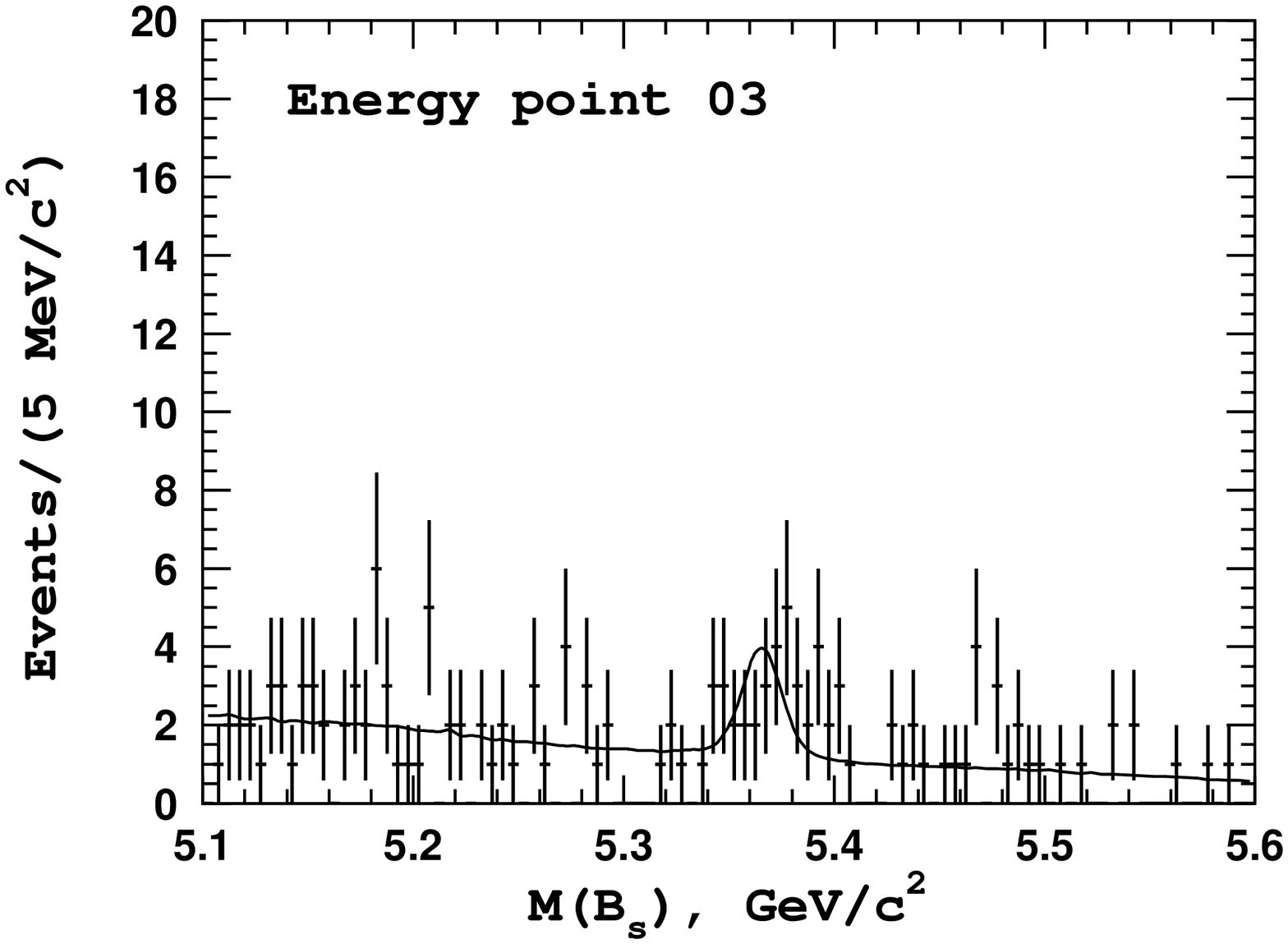} \hfill
  \includegraphics[width=0.24\textwidth]{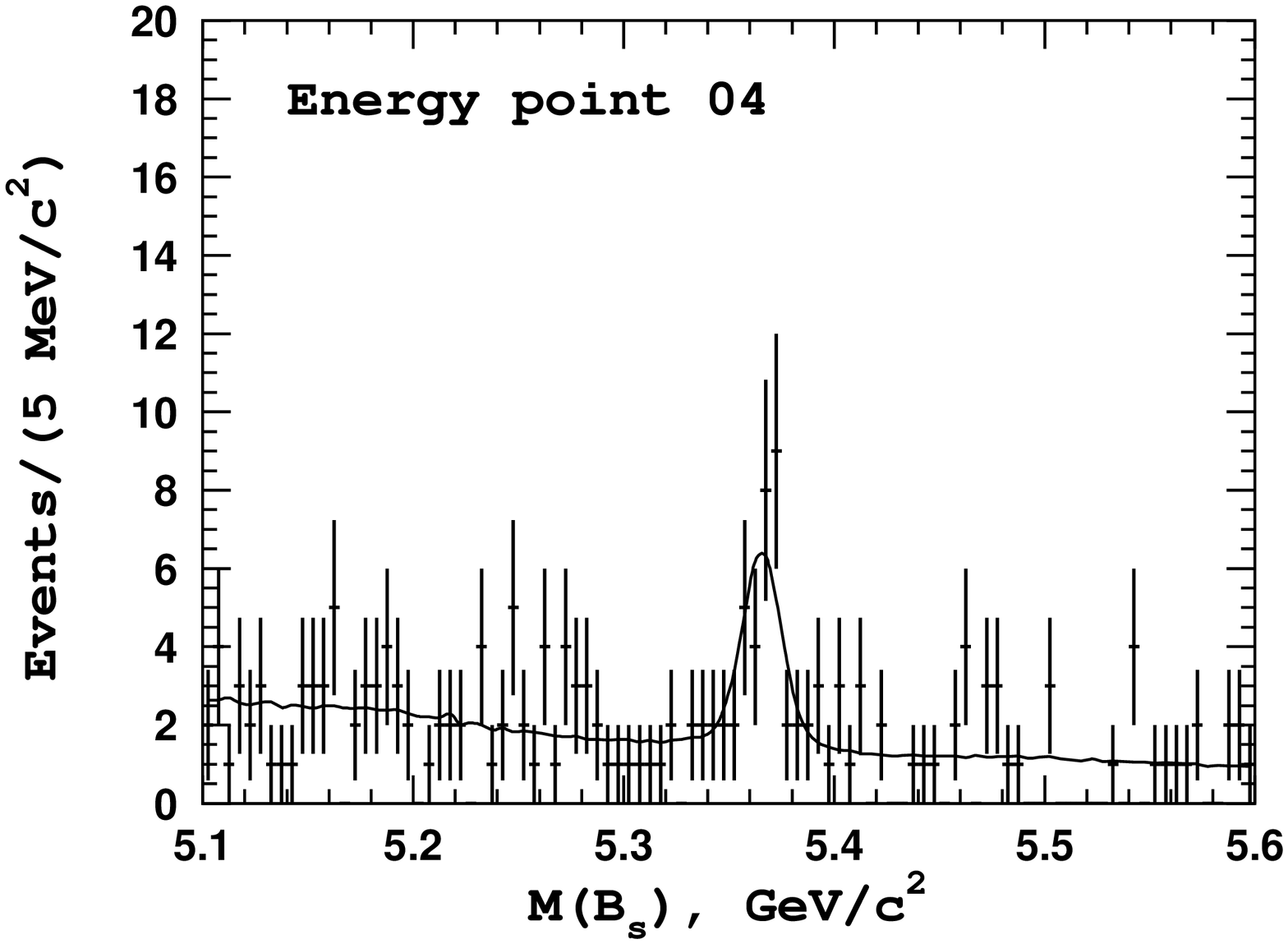} \\
  \includegraphics[width=0.24\textwidth]{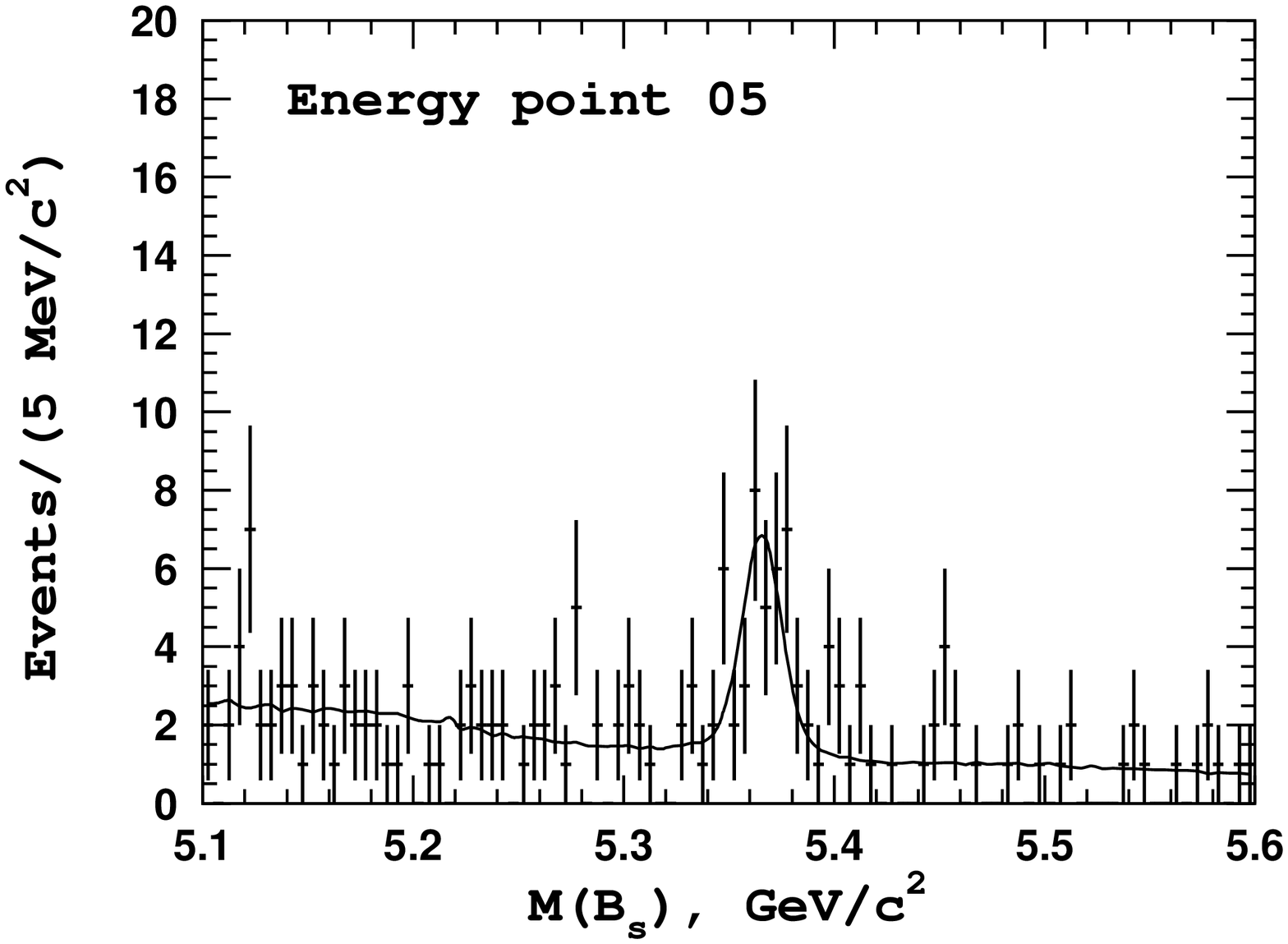} \hfill
  \includegraphics[width=0.24\textwidth]{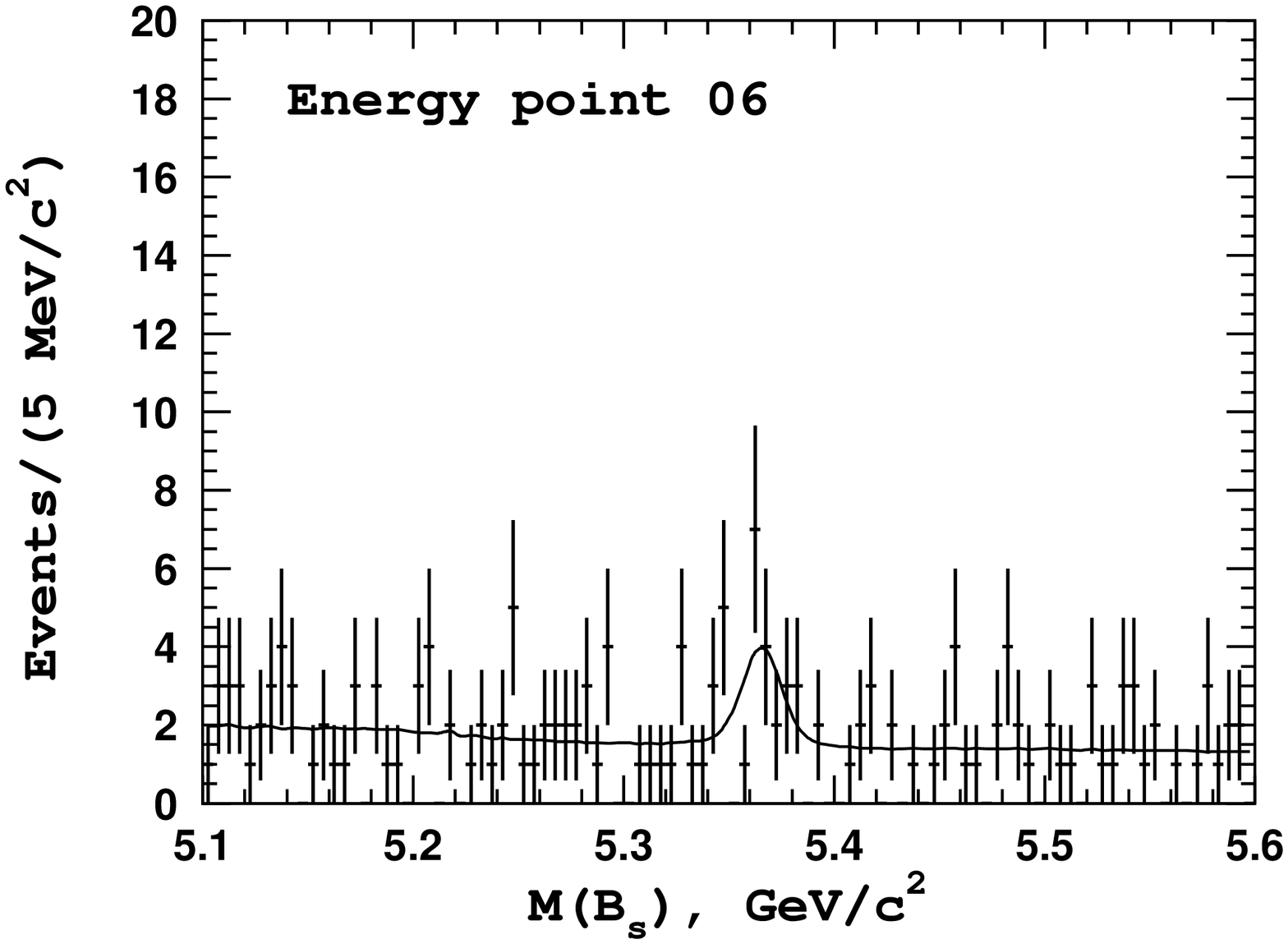} \hfill
  \includegraphics[width=0.24\textwidth]{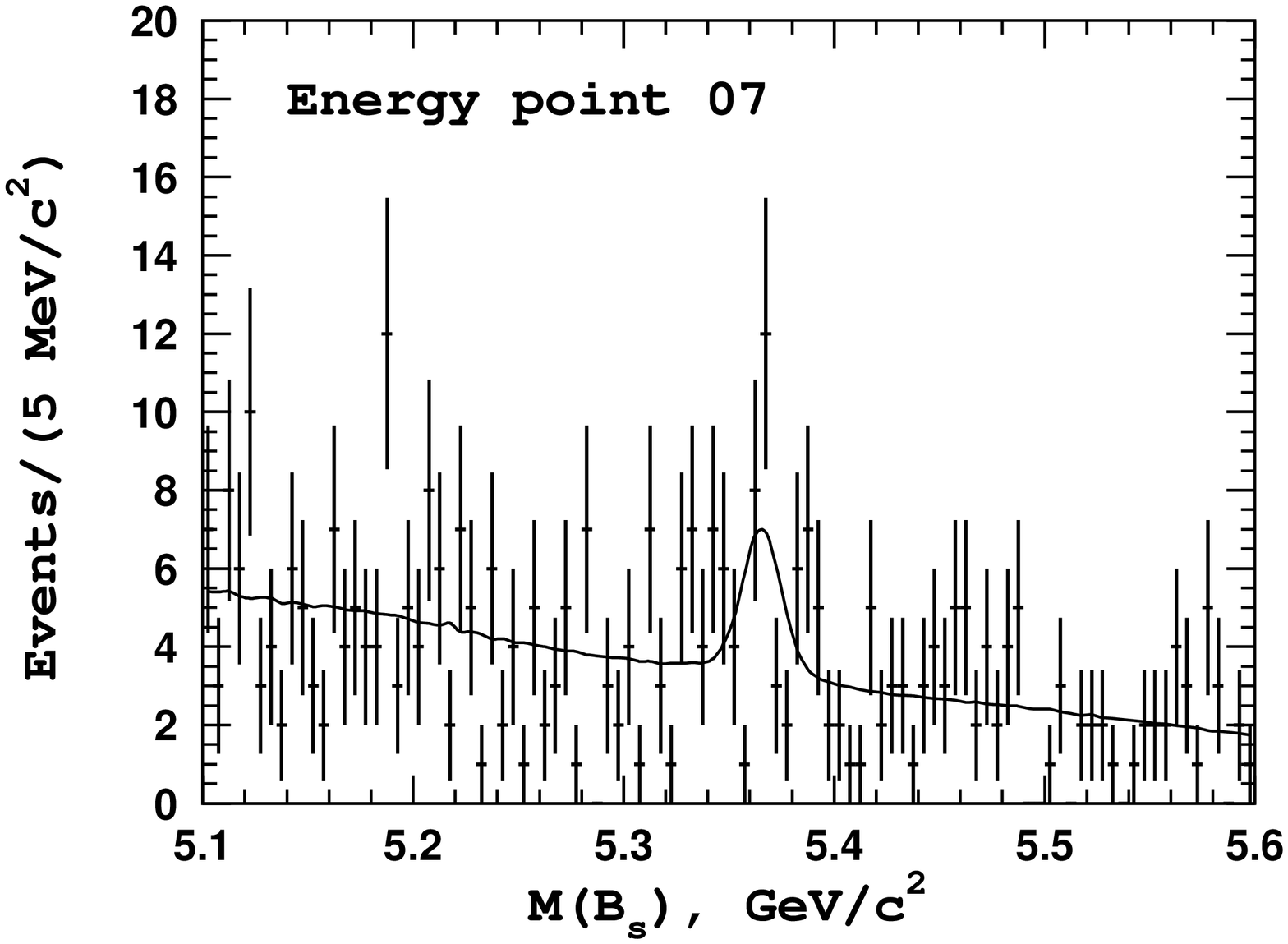} \hfill
  \includegraphics[width=0.24\textwidth]{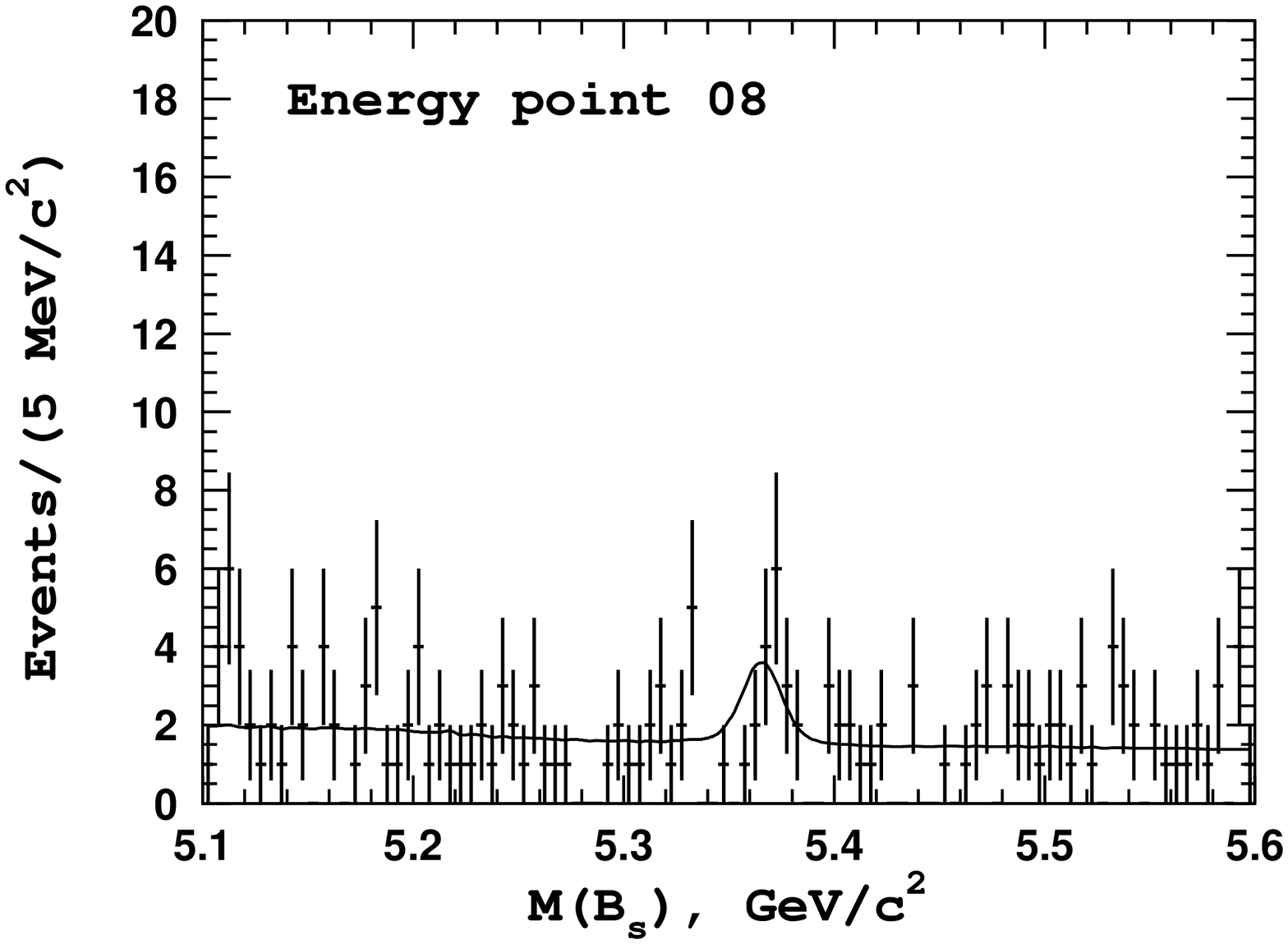} \\
  \includegraphics[width=0.24\textwidth]{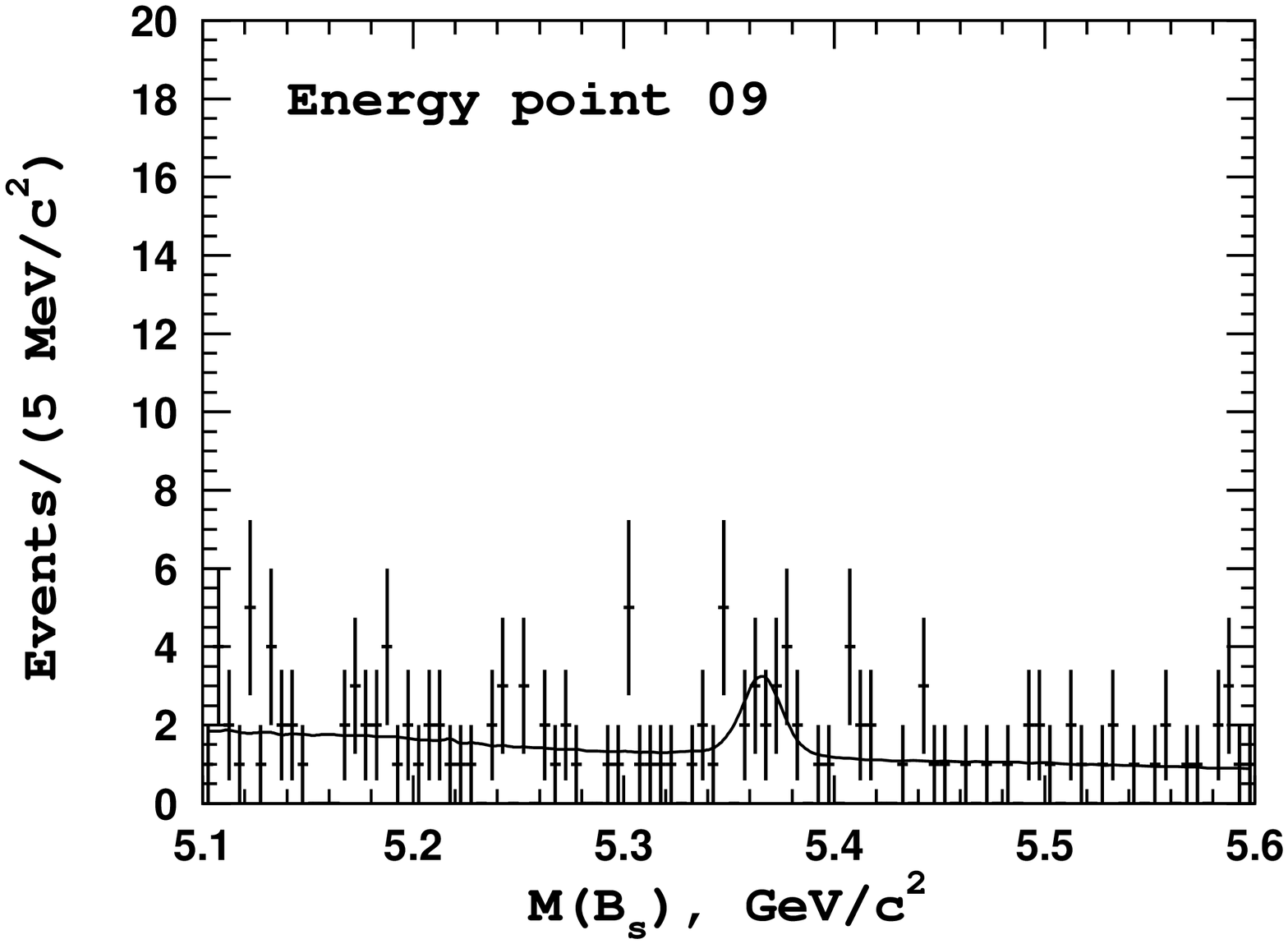} \hfill
  \includegraphics[width=0.24\textwidth]{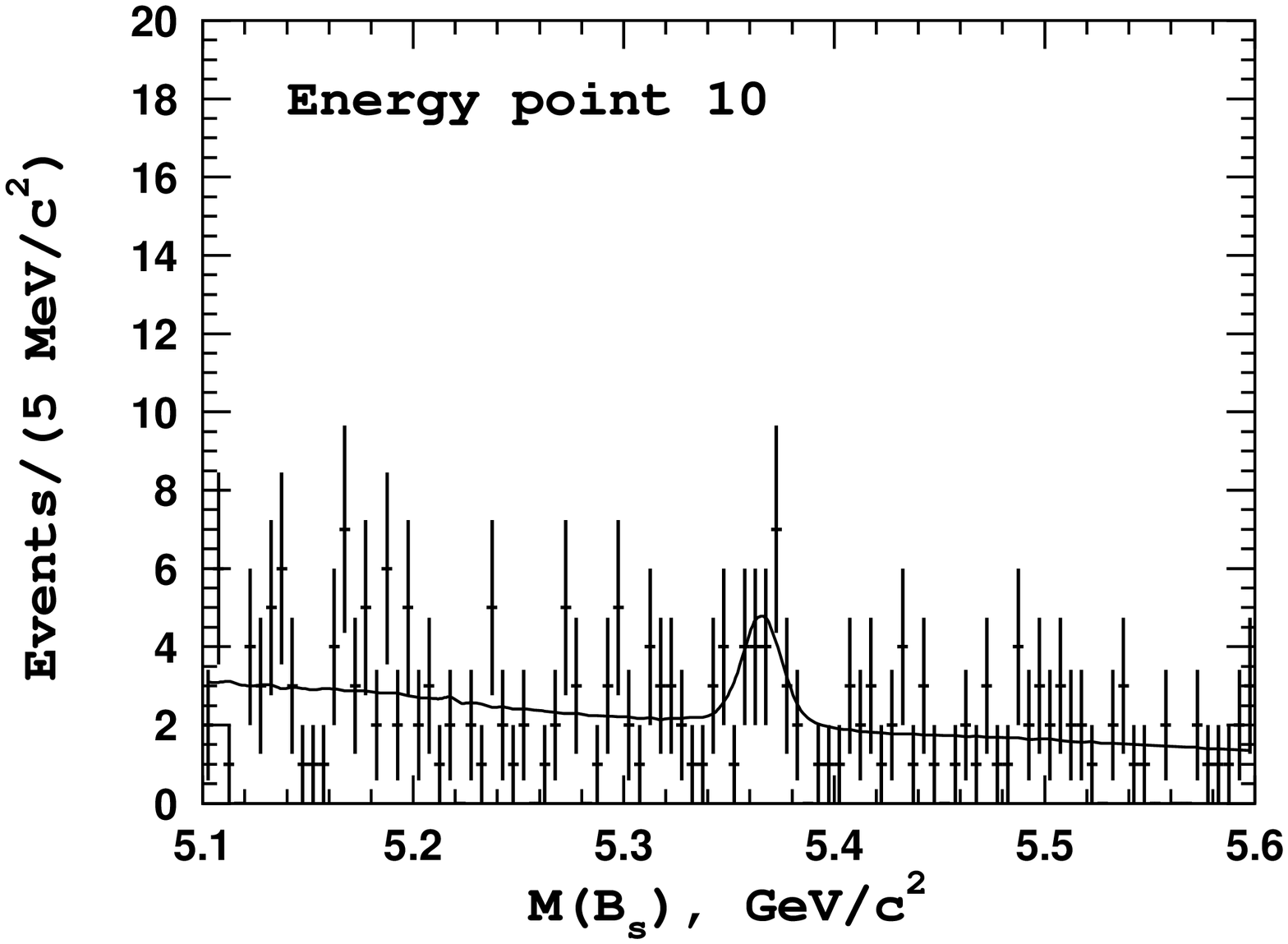} \hfill
  \includegraphics[width=0.24\textwidth]{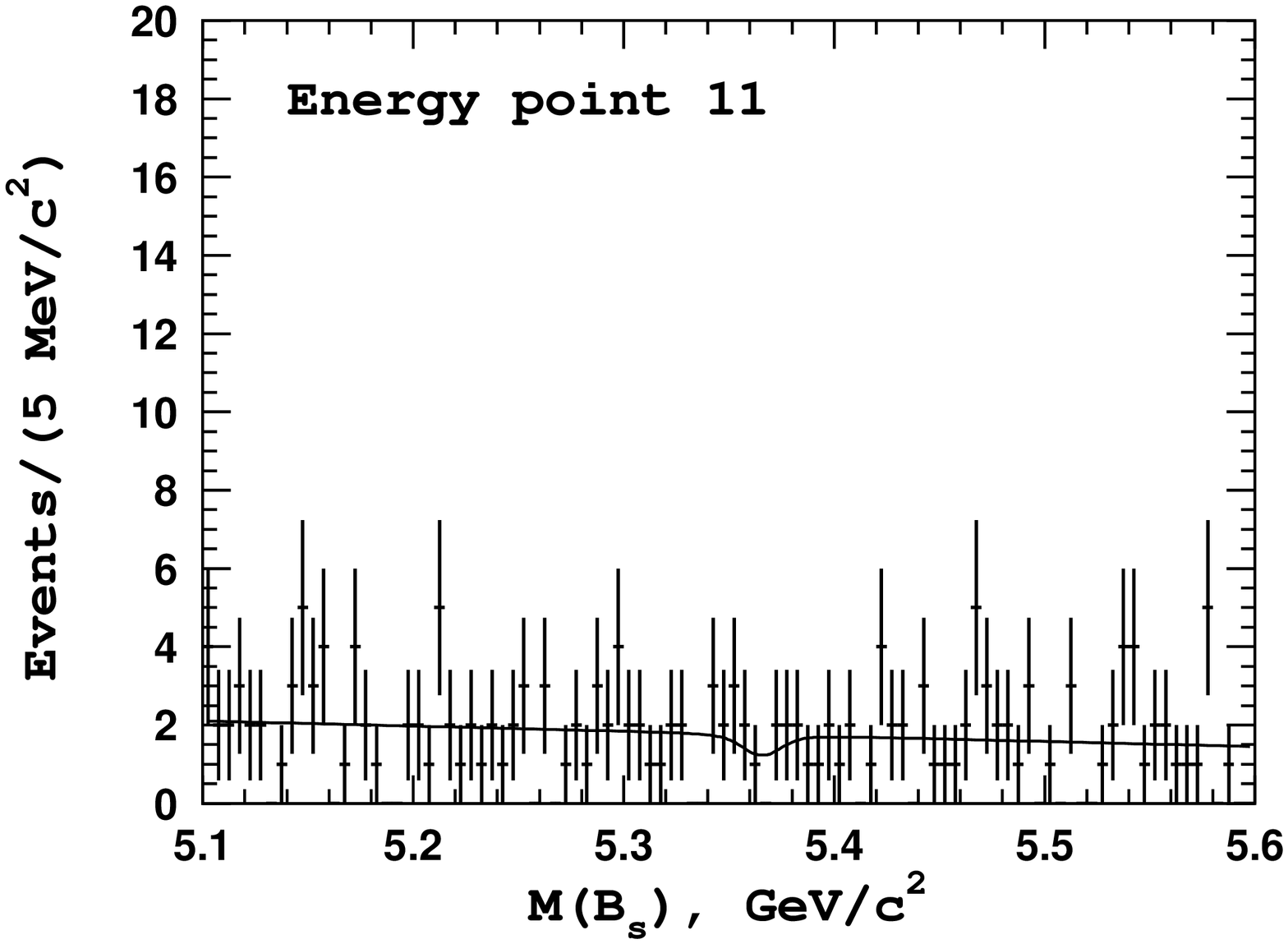} \hfill
  \includegraphics[width=0.24\textwidth]{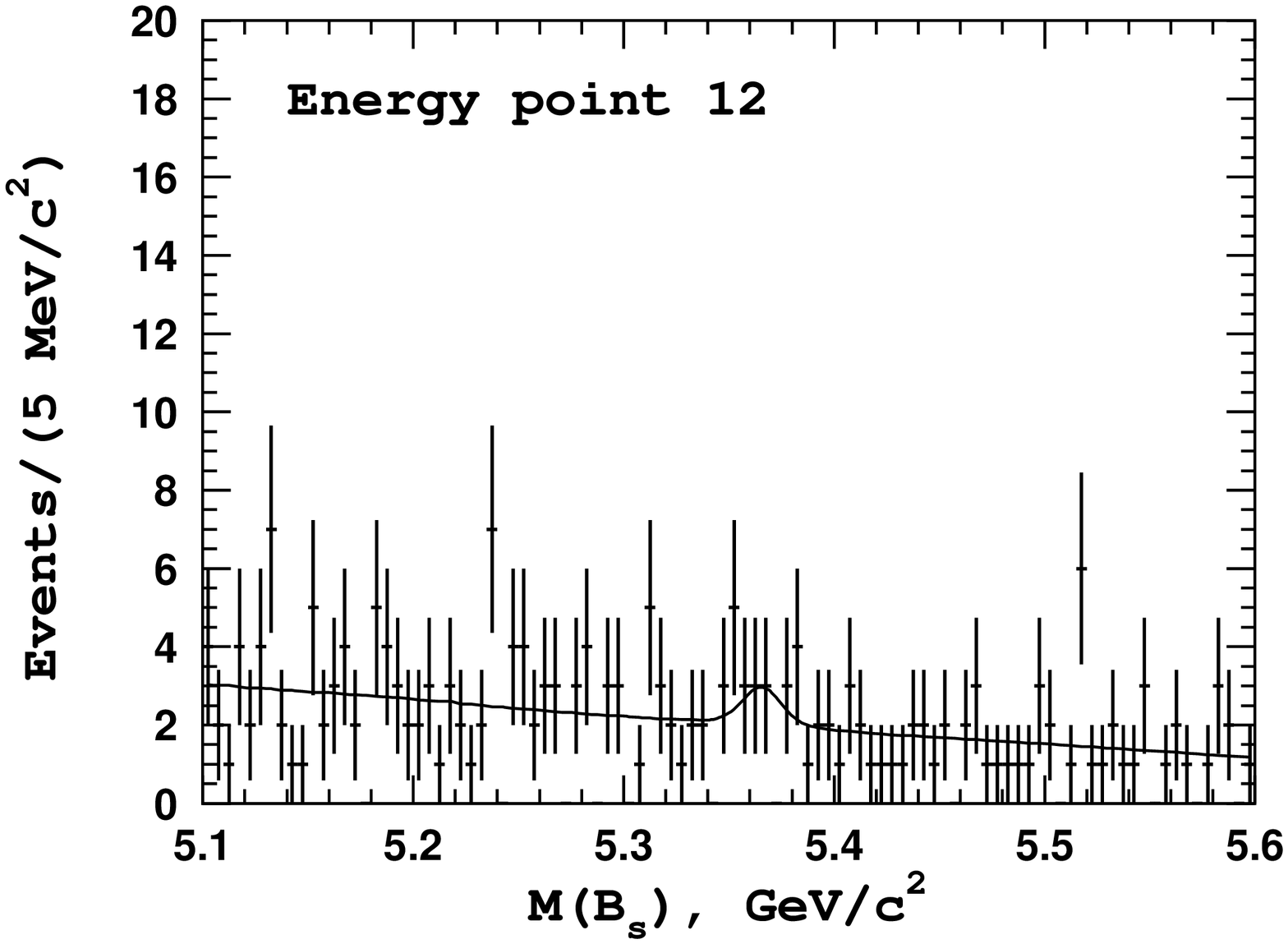} \\
  \includegraphics[width=0.24\textwidth]{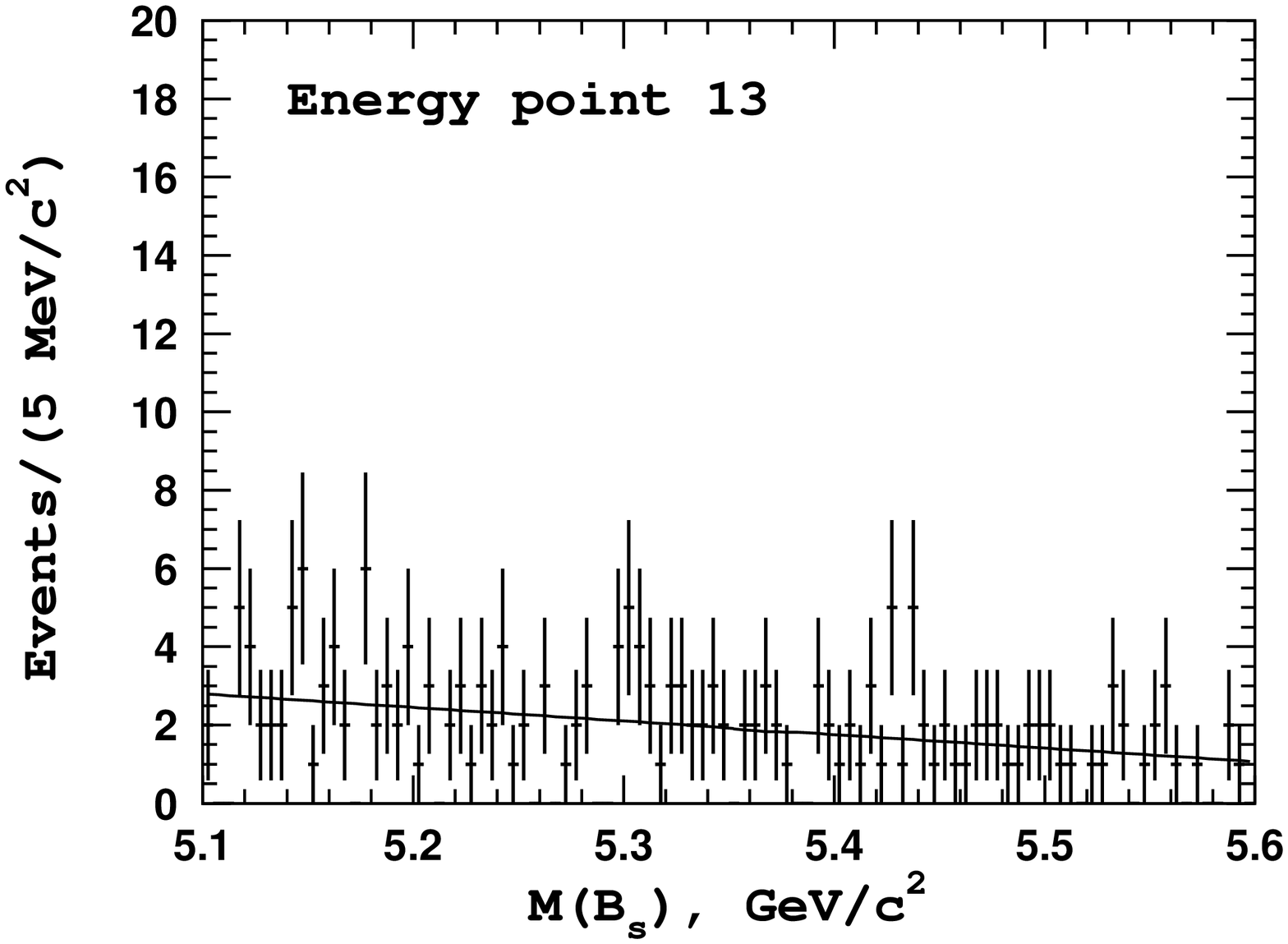} \hfill
  \includegraphics[width=0.24\textwidth]{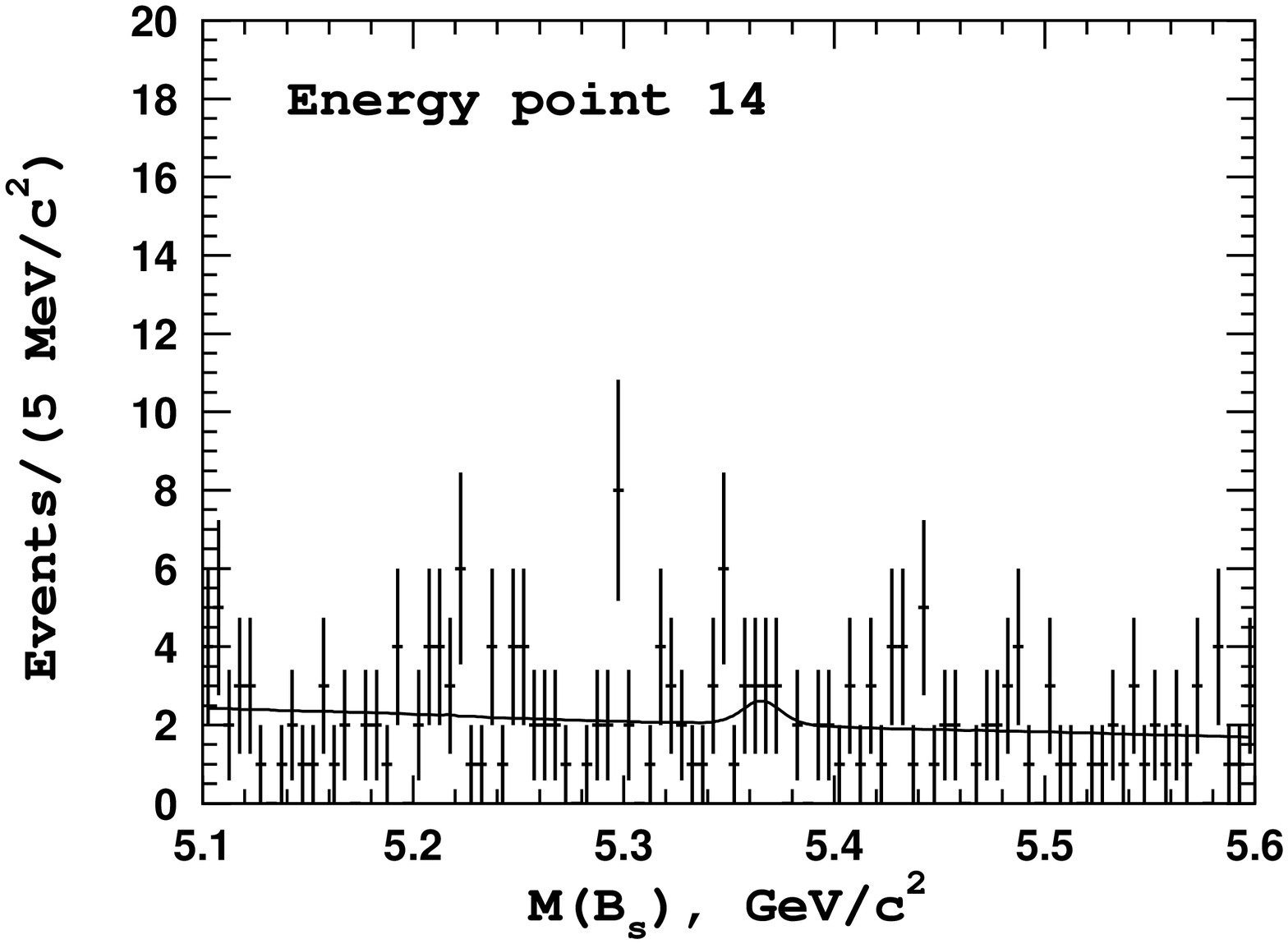} \hfill
  \includegraphics[width=0.24\textwidth]{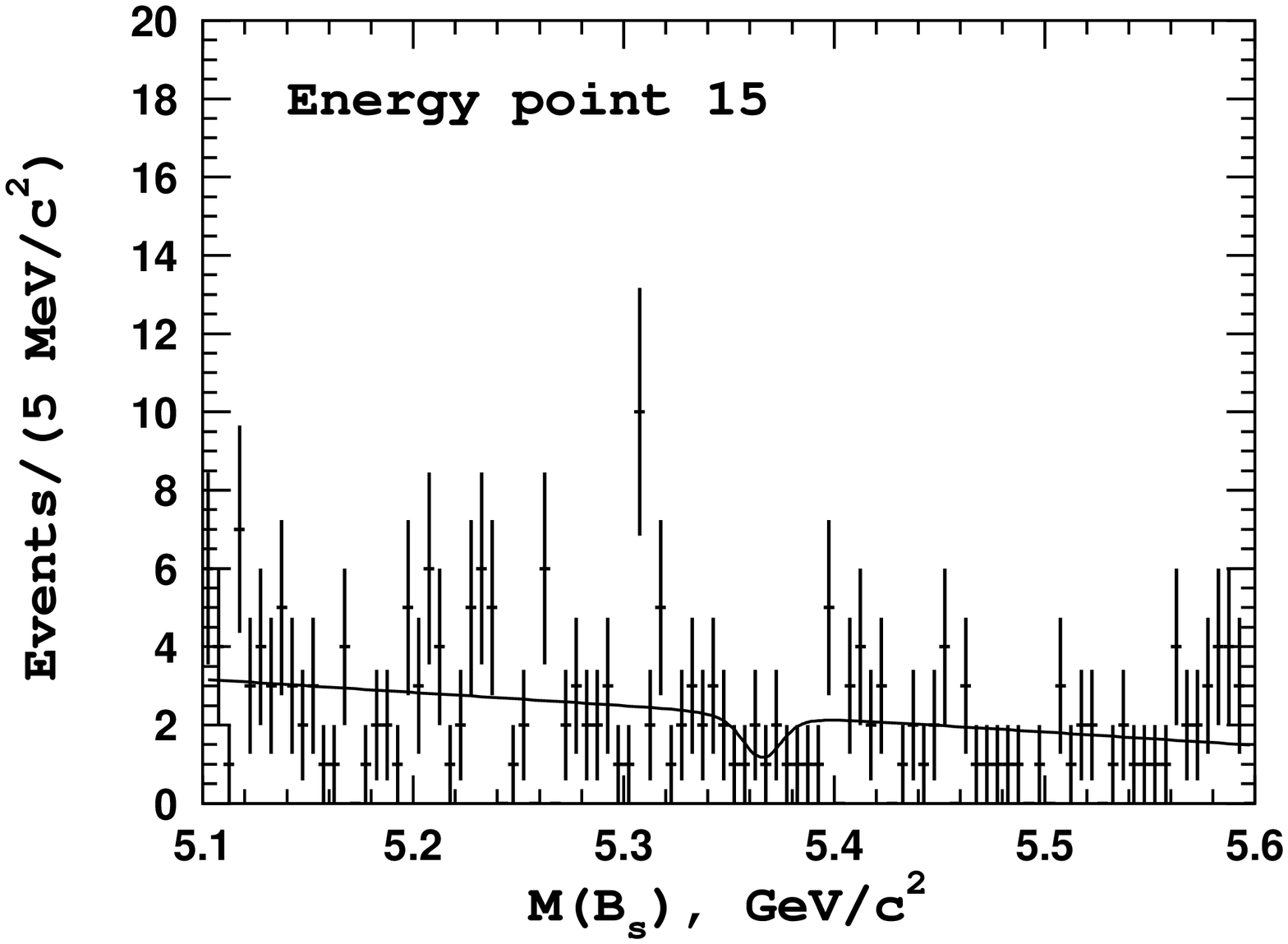} \hfill
  \includegraphics[width=0.24\textwidth]{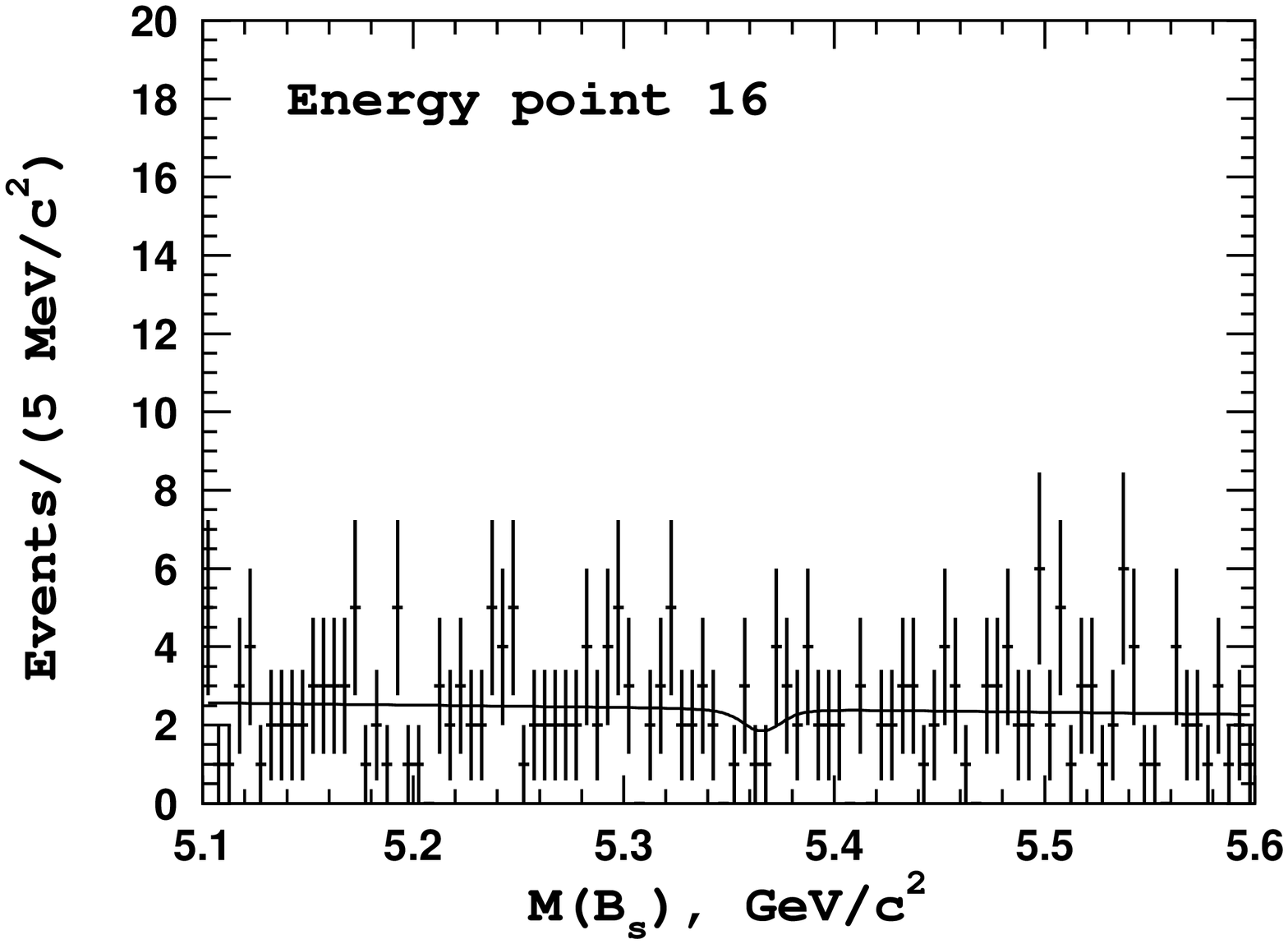} \\
  \includegraphics[width=0.24\textwidth]{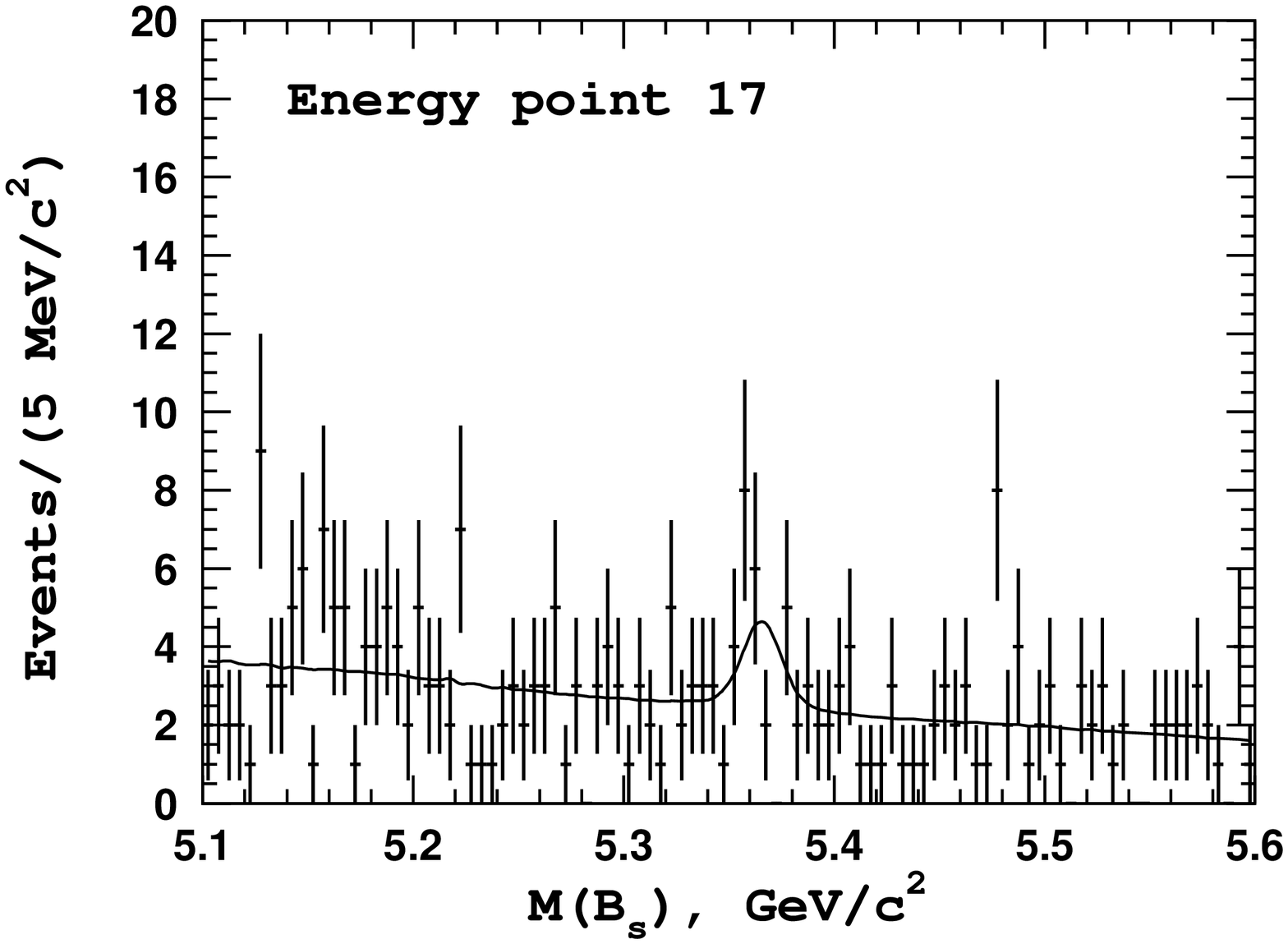} \hfill
  \includegraphics[width=0.24\textwidth]{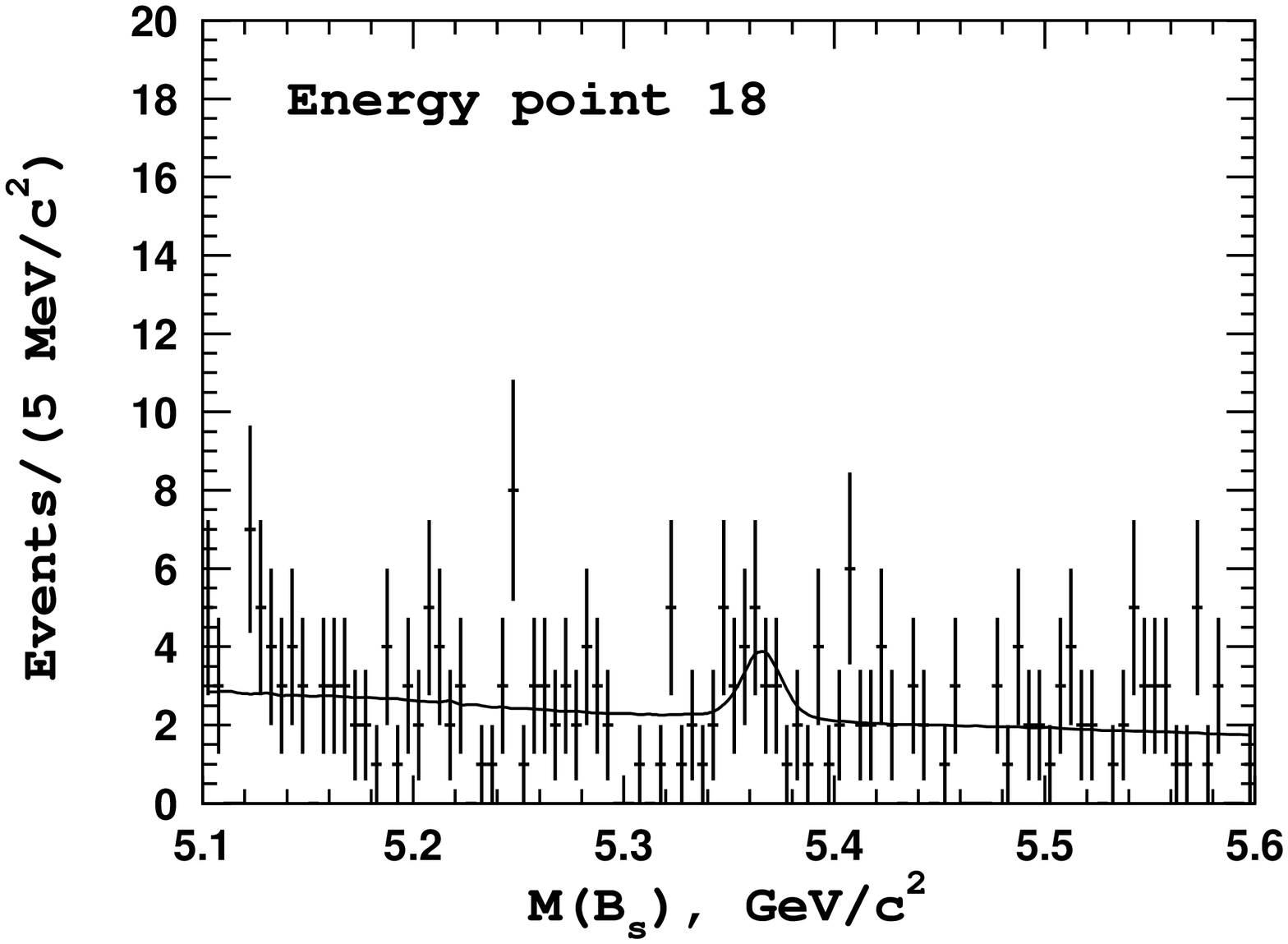} \hfill
  \includegraphics[width=0.24\textwidth]{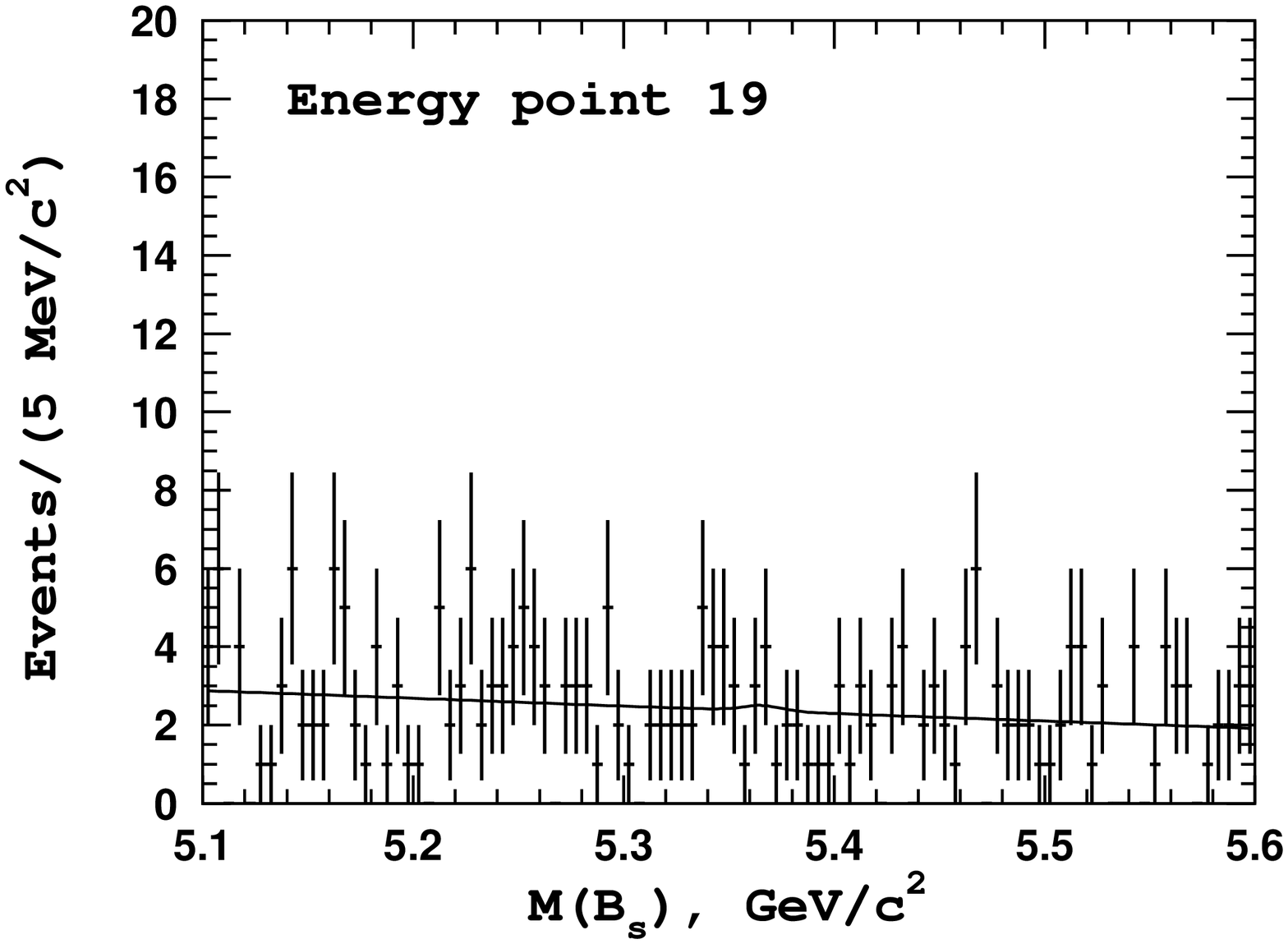} \\
  \includegraphics[width=0.24\textwidth]{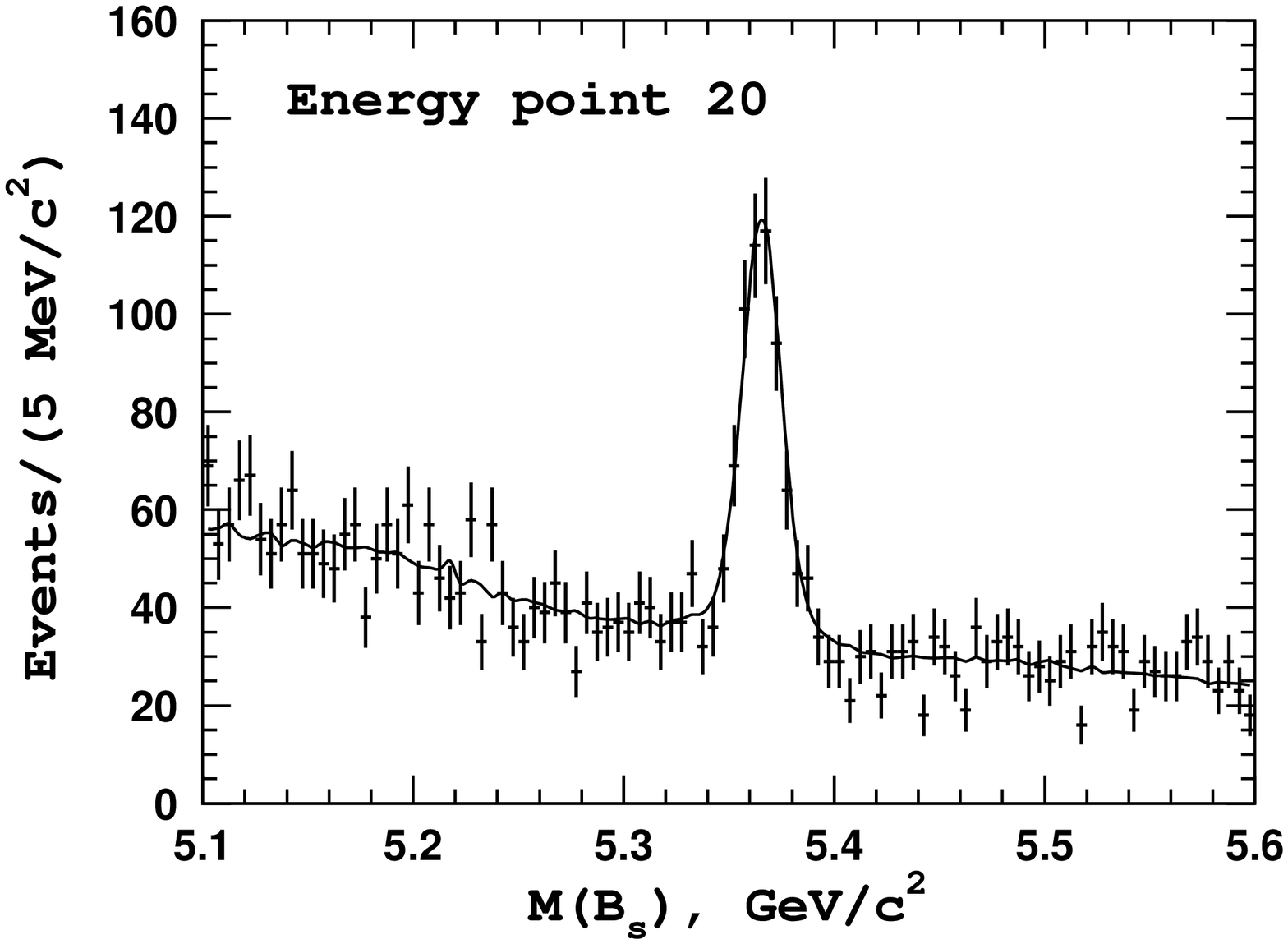} \hfill
  \includegraphics[width=0.24\textwidth]{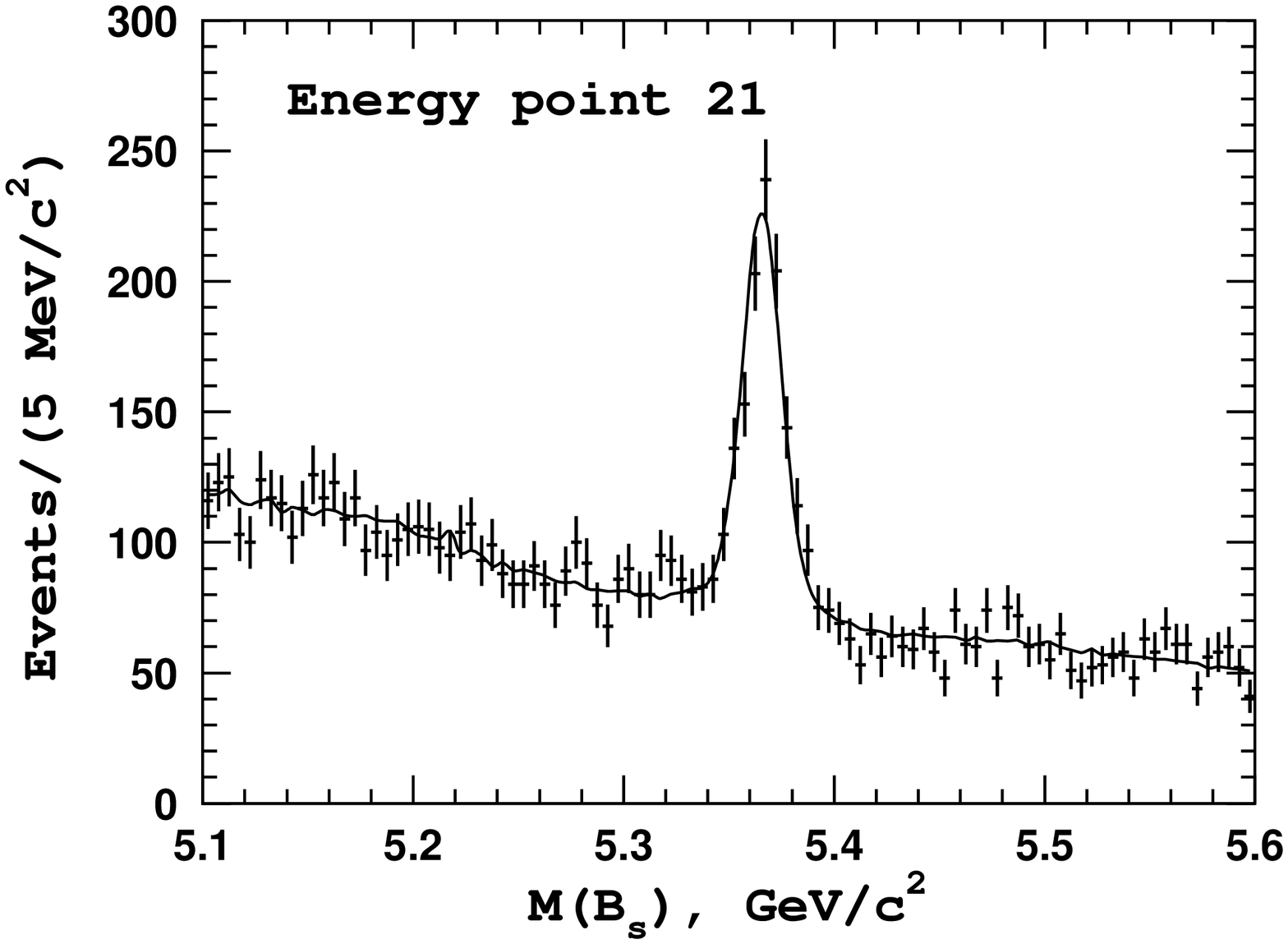} \hfill
  \includegraphics[width=0.24\textwidth]{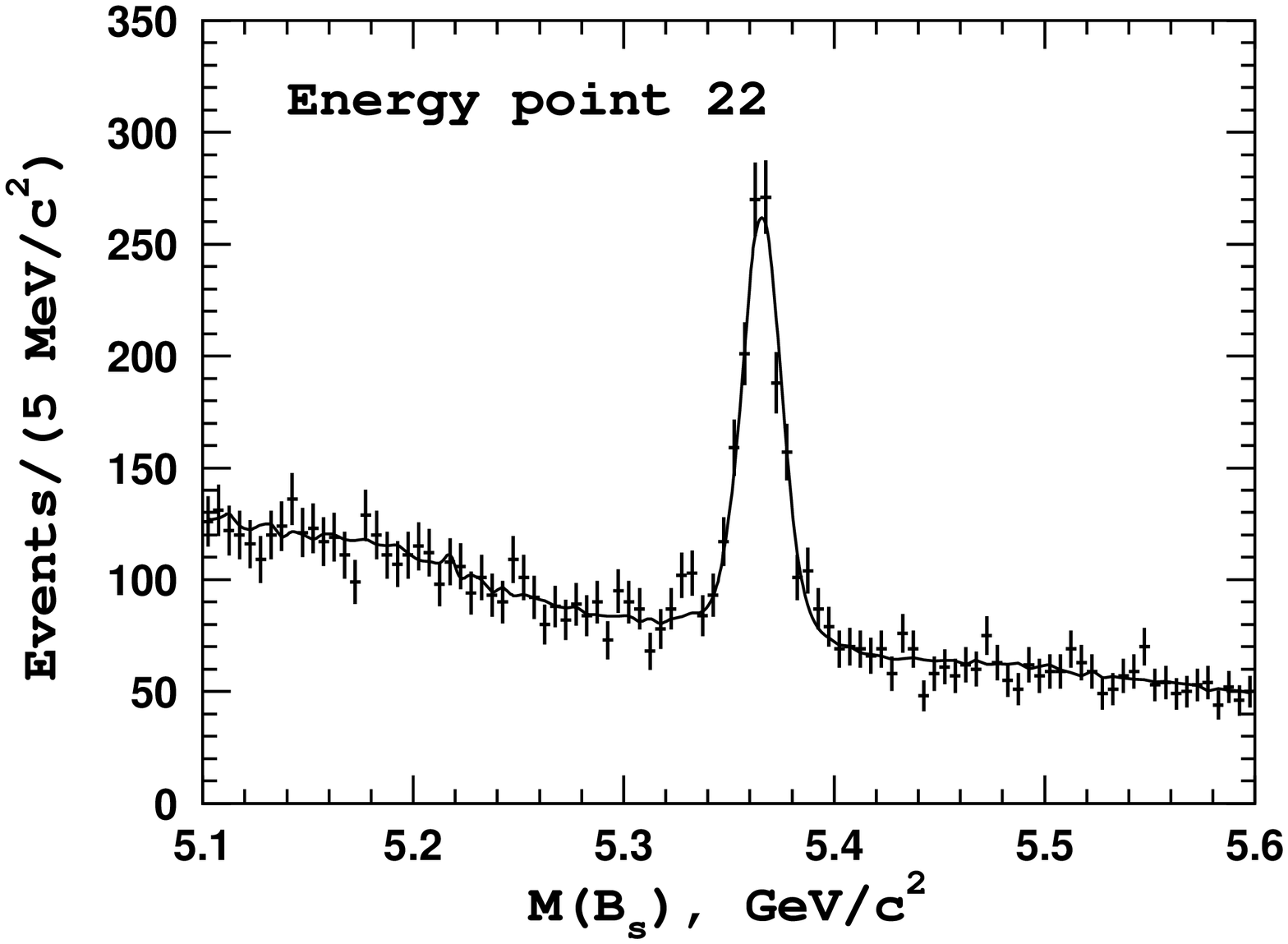} \\
  \caption{$M(B_s)$ distributions for $\ee\to B_s^{(*)}\bar{B}_s^{(*)}$
           candidates for each energy point.}
  \label{fig:e-mass}
\end{figure}

At each energy point, we use the same analysis strategy as applied in 
the analysis of the $\UFS$ data, described in the previous Section. 
The $M(B_s)$ distributions for selected $B_s$ candidates at each energy 
point are shown in Fig~\ref{fig:e-mass}. The relevant information is 
summarized in Table~\ref{tab:e-scan}. 

\begin{figure}[!t]
  \includegraphics[width=0.325\textwidth]{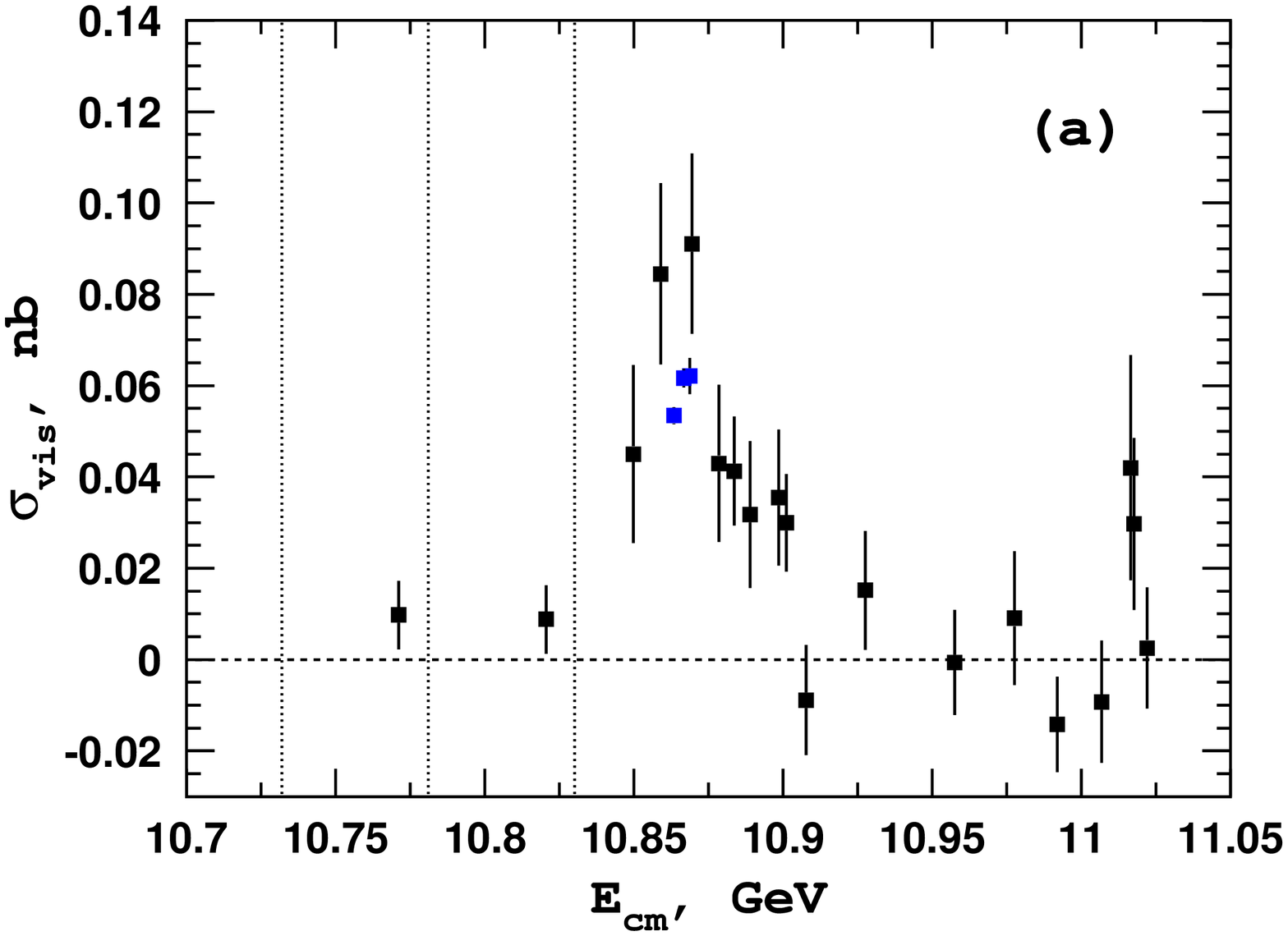} \hfill
  \includegraphics[width=0.325\textwidth]{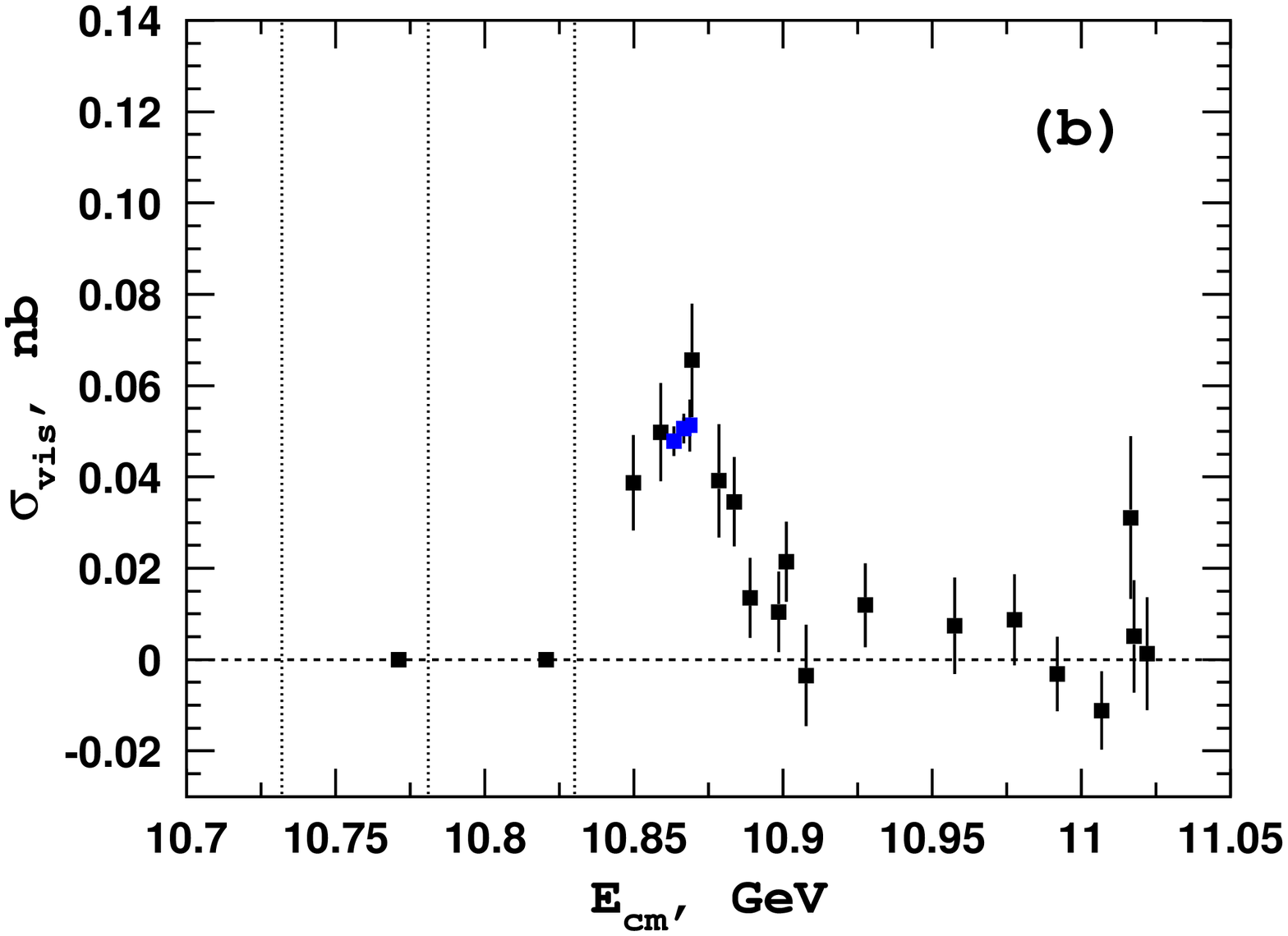} \hfill
  \includegraphics[width=0.325\textwidth]{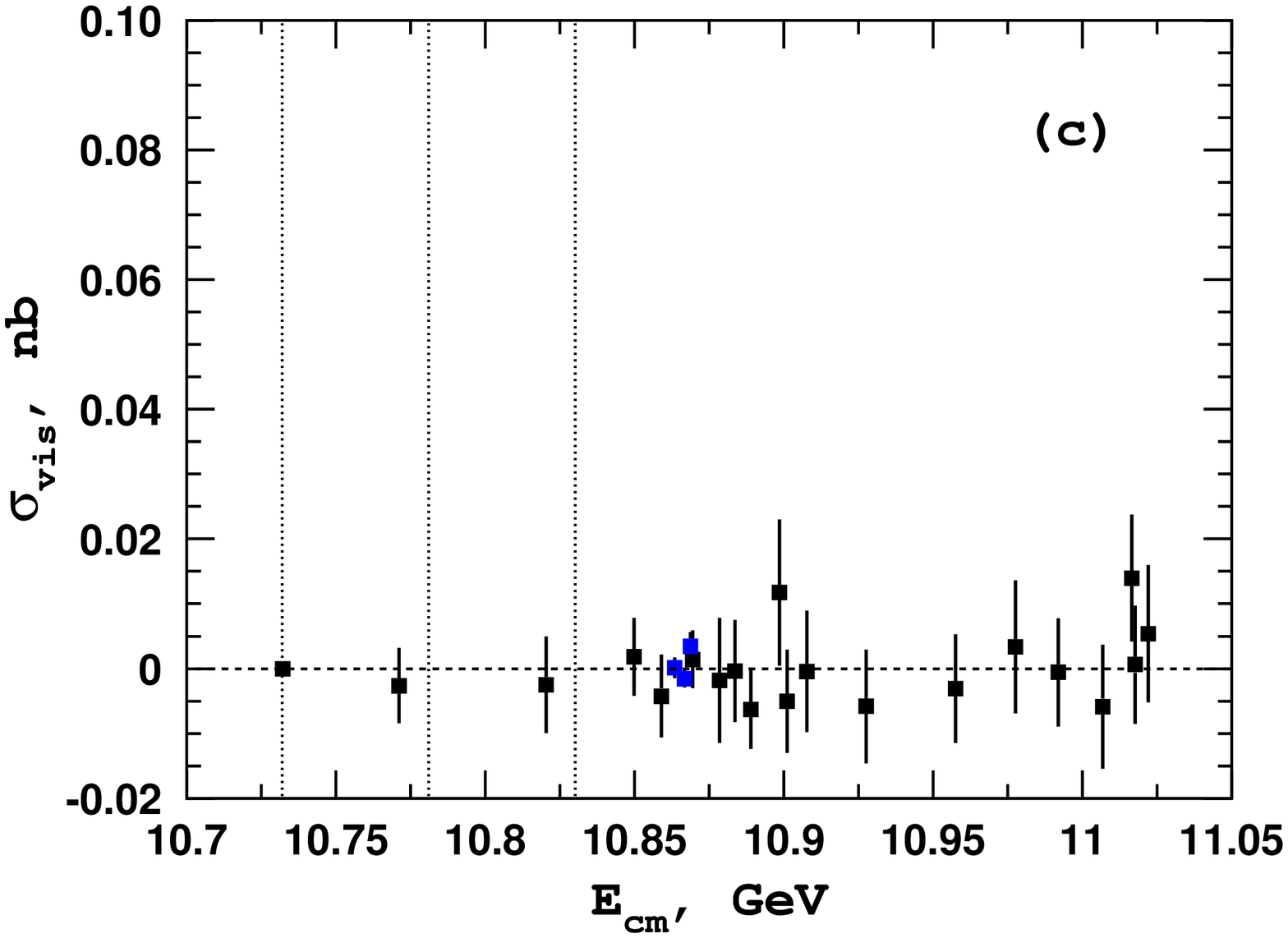} 
  \caption{Cross section for the 
    (a) total $\ee\to B_s^{(*)}\bar{B}_s^{(*)}$;
    (b) $\ee\to B_s^*\bar{B}_s^*$ only;
    (c) momentum sideband region.
    Vertical lines show the $B_s\bar{B}_s$, $B_s\bar{B}_s^*$, and 
    $B_s^*\bar{B}_s^*$ thresholds, respectively.}
  \label{fig:cross}
\end{figure}

The visible cross section $\sigma^{\rm vis}$ shown in 
Fig.~\ref{fig:cross}(a) is calculated as 
\begin{equation}
 \sigma^{\rm vis}_i = 0.0585\frac{N_i}{N_{5S}}\frac{L_{5S}}{L_i},
\end{equation}
where $N_i$ and $N_{5S}=2270\pm60$ are the $B_s$ yields measured at the 
$i$-th energy point and for the full $\UFS$ sample, respectively; $L_i$ 
and $L_{5S}=121.4$~fb$^{-1}$ are the corresponding luminosities. 
The factor $(0.0585\pm0.0106)$~nb is the product of the total 
$\ee\to b\bar{b}$ cross section of $0.340\pm0.016$~nb~\cite{bbcs} and 
the fraction of $\ee\to b\bar{b}$ events hadronized to a pair of $B_s^{(*)}$ 
mesons, measured to be $f_s=0.172\pm0.030$~\cite{bbcs}. Both these 
quantities have been measured by Belle at the $\UFS$.

In addition to the total $\ee\to B_s^{(*)}\bar{B}_s^{(*)}$ cross 
section, we also perform a separate measurement of the exclusive 
$\ee\to B_s^*\bar{B}_s^*$ cross section. We select $B_s^*\bar{B}_s^*$ 
events by applying a tighter requirement on the momentum of the 
reconstructed $B_s$, as summarized in Table~\ref{tab:e-scan}.
Results are presented in 
Fig.~\ref{fig:cross}(b) and in Table~\ref{tab:e-scan}. As a cross 
check, we apply the same procedure to events selected in a 
0.25~GeV/$c$-wide momentum window above the two-body kinematic 
limit. The fit returns a $B_s$ yield consistent with zero at each 
energy point; the measured visible cross section for this sideband 
region is shown in Fig.~\ref{fig:cross}(c).

The systematic uncertainty for the measured visible cross sections 
quoted in Table~\ref{tab:e-scan} is dominated by the common 
multiplicative part due to the uncertainties in the total 
$\ee\to b\bar{b}$ cross section and the hadronization fraction $f_s$. 
The systematic uncertainty due to the $B_s$ signal yield extraction 
is determined for each energy point and varies from 6\% to 20\%.

\section{Conclusion}

In conclusion, the ratio of production cross sections for the two-body 
$ B_s^*\bar{B}_s^*: B_s\bar{B}_s^* +c.c.:B_s\bar{B}_s$ in $\ee$ 
annihilation at $\sqrt{s}=10.866$~GeV is measured to be 
$7:$ $0.853\pm0.106\pm0.053:$ $0.638\pm0.094\pm0.033$. The fraction of 
the $S=0$ component determined from the analysis of the polar angular 
distribution of $B_s^*$ produced in the $\UFS\to B_s^*\bar{B}_s^*$ 
process is $r=0.175\pm0.057^{+0.022}_{-0.018}$.
The measured values of the ratio of the production cross sections and 
fraction of the $S=0$ component are in strong contradiction with the 
HQSS prediction. Some possible reasons for such a difference are 
discussed in Ref.~\cite{Vol2}.
Analysis of the $\UFS\to B_s^*\bar{B}_s^*$ cross section in the energy
range from 10.77 to 11.02~GeV reveals a strong signal of the
$\UFS$ resonance with no statistically significant signal of the
$\USS$ resonance.

\section{acknowledgement}

We thank the KEKB group for excellent operation of the
accelerator;  the  KEK  cryogenics group  for  efficient
solenoid operations; and the KEK computer group, the
NII,  and  PNNL/EMSL  for  valuable  computing  and
SINET4 network support. We acknowledge support from
MEXT, JSPS, and Nagoya’s TLPRC (Japan); ARC and
DIISR (Australia); FWF (Austria); NSFC (China); MSMT
(Czechia); CZF, DFG, and VS (Germany); DST (India);
INFN (Italy); MOE, MSIP, NRF, GSDC of KISTI, and
BK21Plus (Korea); MNiSW and NCN (Poland); MES
(particularly under Contract No. 14.A12.31.0006) and
RFAAE (Russia); ARRS (Slovenia); IKERBASQUE and
UPV/EHU (Spain); SNSF (Switzerland); NSC and MOE
(Taiwan); and DOE and NSF (U.S.).


\end{document}